\documentclass[12pt,a4paper]{article}
\usepackage[T2A,T1]{fontenc}
\usepackage{amssymb,amsmath,stmaryrd,mathtools,bm,graphicx,caption,subcaption,here,parskip,tikz-cd,tikz,ulem,cite,titlesec,enumerate,empheq,changepage,physics,cancel}
\usepackage[rightcaption]{sidecap}
\usepackage[margin=.9in]{geometry}
\usepackage[most]{tcolorbox}
\usepackage[colorlinks,allcolors=blue]{hyperref}

\numberwithin{equation}{section}
\DeclareMathAlphabet{\mathscrbf}{OMS}{mdugm}{b}{n}
\usetikzlibrary{decorations.markings,decorations.pathreplacing}
\newtcbox{\othermathbox}[1][]{nobeforeafter, math upper, tcbox raise base, 
          enhanced, rounded corners, colback=black!5, colframe=black,
          left=0.7em, top=0.4em, right=0.7em, bottom=0.5em}

\definecolor{MyYellow}{RGB}{248,199,82}
\usepackage[utf8]{inputenc} 

\makeatletter
\newcommand{\ostar}{\mathbin{\mathpalette\make@circled\star}}
\newcommand{\make@circled}[2]{%
  \ooalign{$\m@th#1\smallbigcirc{#1}$\cr\hidewidth$\m@th#1#2$\hidewidth\cr}%
}
\newcommand{\smallbigcirc}[1]{%
  \vcenter{\hbox{\scalebox{0.77778}{$\m@th#1\bigcirc$}}}%
}
\makeatother

\renewcommand{\i}[1]{\textit{#1}}

\tikzset{
    matter/.style={thick, line width=1.1pt, draw=black},
    ghost/.style={thick, dashed, line width=1.1pt, draw=black},
    dot/.style={circle, fill=black, minimum size=4.5pt, inner sep=0pt},
    baseline={([yshift=-.5ex]current bounding box.center)}
}

\begin{document}

\begin{center}
{\LARGE{{Quantum Mechanics on Lie Groups:\\[.4em]
II.~Path Integrals}}}\\[1.3em]

{\large Mathieu Beauvillain,$^1$ Blagoje Oblak,$^2$ and Marios Petropoulos$^1$}\\[1.3em]

{\small%
$^1$ CPHT, CNRS, École polytechnique, Institut Polytechnique de Paris, 91120 Palaiseau, France\\
$^2$ Université Claude Bernard Lyon 1, ICJ UMR 5208, CNRS, 69622 Villeurbanne, France}
\end{center}
~\\[2em]

\begin{center}
\begin{minipage}{.8\textwidth}
\textbf{Abstract.} We continue our study of quantum dynamics on a Lie group $G$, initiated in \cite{Beauvillain:2025ygx}, by building the path integral that governs transition amplitudes in the Hilbert space $L^2(G)$. This relies on the proper handling of both a noncommutative momentum space and the presence of compact directions in $G$. We show that compactness can be handled through a sum over winding numbers in maximal tori of $G$, generalizing the similar sum commonly encountered for path integrals on a circle. As applications, we compute semiclassical approximations of propagators and partition functions of Euler-Arnold systems, up to (and including) two-loop order.
\end{minipage}
\end{center}
~\\[2em]
\begin{center}
    Preprint number: CPHT-RR020.072026
\end{center}

\newpage
\tableofcontents

\newpage


\section{Introduction}
\label{sec:intro}

Dynamics on group manifolds is ubiquitous in both classical and quantum physics. Perhaps the oldest nontrivial example is that of rigid bodies or `rotors', whose configuration space is a rotation group \cite{Montgomery,Marsden,holm2024}; but there are many more \cite{ArnoldKhesin,Khesin}, ranging from classical hydrodynamics \cite{ArnoldOrigin,Vasylkevych,Modin,holm1998euler,holm1999euler,gaybalmaz2009reduced,OblakKozy} to quantum liquids \cite{Gripaios:2014yha,Endlich:2010hf,Dersy:2022kjd}, spin chains \cite{LandauLifschitz,Lakshmanan,Jevicki:1978yv,gaudin2014,Lamers,Nussle:2025umz} and condensed matter \cite{Delacretaz,Huang:2023hbt,Beauvillain:2024dou}. The group structure makes such systems highly symmetric, so much so that they often turn out to be classically integrable---although this may not remain true at the quantum level. The simple example of anisotropic rigid bodies falls into that category \cite{Allen,Bauder,Biedenharn:1984,Gripaios:2015pfa}. It seems promising, therefore, to leverage classical information on these systems to study their quantum counterparts. The present paper is part of a series \cite{Beauvillain:2025ygx} aimed at addressing this issue.

We are specifically concerned here with the expression of transition amplitudes on a group $G$ in terms of functional integrals, \i{i.e.}~integrals over paths in phase space $T^*G$, with $G$ interpreted as a `configuration space' and points in cotangent spaces interpreted as `momenta'. To understand the change of basis from position to momentum space, a key preliminary is to build Fourier series for Lie groups that are as general as possible (nonabelian, noncompact). This was addressed in \cite{Beauvillain:2025ygx}, building on earlier works in quantum gravity \cite{Freidel:2005bb,Freidel:2005me,Freidel:2005ec,Joung:2008mr,Dupuis:2011fx,Oriti:2011ac,Raasakka:2011np,Guedes:2013vi}; we refer to ref.~\cite{Beauvillain:2025ygx} for details that will often be omitted below. Equipped with this tool, we derive path integrals through repeated insertion of the identity in the time-evolution kernel, or \i{propagator}, adapting Feynman's textbook derivation to the Lie group setting. Our approach relies on a decompactification of the group on its Lie algebra, similar to what is commonly done for the U(1) case \cite{Schulman}. It should therefore pave the way for a rigorous definition of path integrals on group manifolds.

Having set up path integrals, one is typically interested in their values in the semiclassical limit. This will be our perspective too, focussing in particular on Lie group dynamics whose Hamiltonian is `purely kinetic' in the sense that it only depends on momenta. Such \i{Lie-Poisson systems} enjoy a nonabelian translation symmetry given by the group itself \cite{Marsden,holm2024,ArnoldOrigin,OblakKozy}, and they cover all the examples cited above. In fact, the subclass of these systems that is most relevant for physics is that of \i{Euler-Arnold systems} \cite{Khesin}, whose Hamiltonian is quadratic. They may be seen as nonabelian generalizations of free particles. Indeed, their classical dynamics notoriously reduces to geodesic motion on a Lie group $G$ endowed with a (one-sided) invariant metric \cite{ArnoldOrigin}.

At the quantum level, the Hamiltonian operator of an Euler-Arnold system is an (a\-ni\-so\-tro\-pic) Laplacian on $G$, so the corresponding propagator in $L^2(G)$ reduces to a heat kernel. Heat kernels on symmetric spaces have been studied at length, \cite{Mori:2019,Craddock:2017,David:2010,avramidi:2006,Vassilevich:2003xt}, but results on Lie groups for arbitrary invariant metrics remain scarce. Our approach differs from standard heat kernel computations in several ways. Most notably, the expansion around classical trajectories does not rely on Riemann normal coordinates, but rather on the Lie group structure. This has the advantage of yielding an extensive list of vertices to be used at all loop orders. Another subtlety to keep in mind is that quantum Euler-Arnold systems are generally neither integrable, nor are their path integrals one-loop exact. Indeed, our results include explicit two-loop expansions for both transition amplitudes and partition functions of Euler-Arnold systems.

Previous attempts to derive Lie group path integrals from noncommutative Fourier transforms emanated from loop quantum gravity \cite{Freidel:2005bb,Freidel:2005me,Freidel:2005ec,Raasakka:2011np} (see also \cite{kapranov}). However, the framework proposed in that context is incomplete, as it fails to define Fourier coefficients and Fourier series when the group $G$ admits compact directions \cite{Beauvillain:2025ygx}. We show instead that the proper treatment of compact subgroups ultimately reproduces the correct sum over instanton sectors, generalizing the sum over winding numbers that appears in the U(1) case. Moreover, refs.~\cite{Freidel:2005bb,Freidel:2005me,Freidel:2005ec,Joung:2008mr,Dupuis:2011fx,Oriti:2011ac,Raasakka:2011np,Guedes:2013vi} overlook the case of non-unimodular groups, which we treat with extra care. We therefore always distinguish left and right Haar measures on a Lie group, which entails the presence of modular functions in many of our formulas.\footnote{`Modular function' is meant here in the sense of measure theory---nothing to do with complex analysis.}

The article is organized as follows. Since our aim is semiclassical analysis, we first review classical motion on Lie groups in section \ref{sec:classical}. In section \ref{sec:review}, we recall the tools defined in \cite{Beauvillain:2025ygx}, with extra emphasis on the Dirac notation in a Lie group setting. This is used in section \ref{sec:G} to write the path-integral representation of a fully generic propagator on a Lie group. We recommend that readers unfamiliar with the U(1) path integral (or the compactified boson) read appendix~\ref{app:U1} before diving into section \ref{sec:G}. Given the physical relevance of Euler-Arnold systems, we focus on them in section \ref{sec:G2} and give a clean recipe to compute the corresponding path integral at the desired loop order. We carry this out explicitly up to (and including) two loops. One may also be interested in equilibrium properties of quantum Euler-Arnold systems; these are captured by canonical partition functions, studied in section \ref{sec:partition}. 
Various technical details are deferred to appendices \ref{app:B}--\ref{app:2loop}.


\section{Classical mechanics on Lie groups}
\label{sec:classical}

The main use of path integral quantization is to extract information about a quantum system by perturbing around classical trajectories. This section is therefore devoted to classical motion on Lie groups \cite{Marsden,holm2024,Khesin}. We first review the Hamiltonian formalism, then show how it maps to a Lagrangian variational problem.

\subsection{Hamiltonian Lie-Poisson systems}
\label{ssec:hamilt}

The Hamiltonian formalism is the framework that allows for canonical quantization, naturally leading to the definition of a Hilbert space acted upon by an operator algebra. To set the stage for quantization in section \ref{sec:review}, we recall the definition of Poisson brackets for dynamics on Lie groups and derive the corresponding equations of motion---first for a generic Hamiltonian, then for the special case of Lie-Poisson and Euler-Arnold Hamiltonians. 

\paragraph{Poisson brackets.} We use the same notation as in \cite[sec.~2]{Beauvillain:2025ygx}: our configuration space is a Lie group $G$ (with elements $g,h$, identity $e$, dimension $n$) whose Lie algebra $\mathfrak{g}$ (with elements $X,Y$) is viewed as the space of right-invariant vector fields on $G$.\footnote{Lie algebras are normally defined by left-invariant vector fields, but we choose right-invariant vector fields to ensure that they generate the \i{left} regular representation: see eq.~\eqref{lereg}.} The corresponding phase space $T^*G\cong G\times\mathfrak{g}^*$ is a trivial bundle (with `momenta' $p,q\in\mathfrak{g}^*$). Its symplectic form can be written as $-\dd\mathcal A$ in terms of the Liouville one-form (see \i{e.g.}~\cite[app.~A]{OblakKozy})
\begin{align}
    \mathcal A_{(g,p)} 
    :=\langle p, \dd g g^{-1}\rangle
    =
    \langle p, \Omega^{\text{R}}_g\rangle,
    \label{liouville}
\end{align}
where $\Omega^{\text{R}}_g:=\dd g g^{-1}$ is the right Maurer-Cartan form at $g\in G$ and $\langle \cdot,\cdot\rangle$ denotes the pairing between $\mathfrak{g}^*$ and $\mathfrak{g}$.

To define coordinates, let $\{t_i|i=1,\dots,n\}$ be a basis of $\mathfrak{g}$ such that
\begin{equation}
\label{struc}
[t_i,t_j]=c{^k}{_{ij}}t_k
\end{equation}
for some structure constants $c_{ij}{^k}$ (with implicit summation over repeated indices). Also let $\{(t^i)^*|i=1,\dots, n\}$ be the corresponding dual basis of $\mathfrak{g}^*$, such that $\langle(t^i)^*,t_j\rangle=\delta^i_j$, and write any momentum as $p=p_i(t^i)^*\in\mathfrak{g}^*$. Then the Poisson bracket induced by the symplectic form defined above \eqref{liouville} is
\begin{align}
      \{A,B\}
      =
      \mathcal{L}_i A\frac{\partial B}{\partial p_i} -
      \frac{\partial A}{\partial p_i}\mathcal{L}_i B - c^k{}_{ij} \frac{\partial A}{\partial p_i}\frac{\partial B}{\partial p_j}p_k
      \label{e22}
\end{align}
for any two functions $A(g,p)$ and $B(g,p)$, where ${\cal L}_i:={\cal L}_{t_i}$ is the Lie derivative along $t_i$. Taking $B=p_j$ in \eqref{e22} shows that momentum coordinates are the Hamiltonian generators of left translations when acting on functions on $G$; this will translate into the left regular representation in section \ref{sec:review}.

\paragraph{Time evolution.} Let $H(g,p)$ be a Hamiltonian function on $G\times\mathfrak{g}^*$. The corresponding equations of motion follow from the Poisson brackets \eqref{e22}, and read
\begin{align}
\label{3a}
      \dot gg ^{-1}\,  &= \partial_p H,
      \\
      \dot p - \text{ad}^*_{\dot g g ^{-1}}p &= - R_g^* \partial_g H,
      \label{3b}
\end{align}
where $(g(t),p(t))$ is a path in $G\times\mathfrak{g}^*$ and dots denote  time derivatives. On the right-hand side, $\partial_pH$ and $\partial_gH$ are the exterior derivatives of the functions $H(g,\cdot)$ and $H(\cdot,p)$ on $\mathfrak{g}^*$ and $G$, respectively. What we write as $R_g^*$ is the pullback, to the identity, by right multiplication $R_g:h\mapsto hg$. On the left-hand side of \eqref{3b}, $\text{ad}^*$ is the coadjoint representation of $\mathfrak{g}$, defined by $\langle\text{ad}^*_X(p),Y\rangle:=-\langle p,[X,Y]\rangle$ for any $X,Y\in\mathfrak{g}$ and any $p\in\mathfrak{g}^*$.

In the special case where $H(g,p)=H(p)$ only depends on momenta, eq.~\eqref{3b} decouples from the `position' dynamics of $g(t)$, yielding the \i{Lie-Poisson equation} \cite{Vasylkevych,OblakKozy}
\begin{align}
    \dot p = \text{ad}^*_{\partial_p H}p.
    \label{e27}
\end{align}
This is a first-order evolution equation for $p(t)$; its solution is unique, given an initial momentum $p(0)$. The time-dependent position $g(t)$ can then be inferred by integrating \eqref{3a} from an initial position $g(0)$---a process sometimes called `reconstruction' \cite{Montgomery,Marsden}. In fact, the whole motion of $p(t)$ given by \eqref{e27} is localized on a coadjoint orbit of $G$, as the charge (`momentum map')
\begin{align}
    Q = \text{Ad}^*_{g^{-1}}(p)
    \label{momentmap}
\end{align}
is conserved in time by virtue of eqs.~\eqref{3a}--\eqref{3b} with $\partial_gH=0$. The reduced phase space for Lie-Poisson dynamics is thus a coadjoint orbit of $G$.

The Hilbert space given by geometric quantization of the full phase space $T^*G$ is $L^2(G)$, \i{i.e.}~the space of square-integrable wave functions on $G$ \cite{woodhouse1997}. By contrast, the reduced phase space (a coadjoint orbit) yields the smaller Hilbert space of a single irreducible unitary representation of $G$. A telling example to keep in mind is that of rotational dynamics in $G=\text{SO}(3)$, for which $L^2(G)$ describes, say, the orbital degrees of freedom of a molecule \cite{Allen,Bauder}, while a quantized coadjoint orbit describes a single spin degree of freedom. Our approach here will consist in defining path integrals for the larger Hilbert space $L^2(G)$, rather than (possibly more standard) coherent-state path integrals in the carrier space of an irreducible representation of $G$.

\paragraph{Euler-Arnold dynamics.} The most ubiquitous class of Lie-Poisson dynamics is that of \i{Euler-Arnold} systems \cite{Khesin}. These correspond to the even more restricted case where $H(p)$ is a quadratic function of momenta:
\begin{align}
     H
     =
      \frac{1}{2}\big\langle p, I^{-1}(p)\big\rangle,
     \label{euler-arnold}
\end{align}
where the \i{inertia tensor} $I:\mathfrak g\rightarrow \mathfrak g^*$ is an invertible linear map that is symmetric in the sense that $\langle I X,Y\rangle = \langle IY,X\rangle$. In that case, $\partial_p H = I^{-1}(p)$, so eq.~\eqref{3a} linking momenta and velocities reads
\begin{align}
    p=I(\dot g g^{-1}),
    \label{velmom}
\end{align}
and the Lie-Poisson equation \eqref{e27} becomes
\begin{align}
    \dot p = \text{ad}^*_{I^{-1}(p)}p.
    \label{euler_arnold_eom}
\end{align}
Most of the present work will be concerned with such Euler-Arnold systems. We stress that the quadratic structure implies neither one-loop exactness nor integrability in quantum theory, due to the nonabelian nature of the group $G$.

\subsection{Lagrangian approach; geodesic motion}
\label{sselag}

While the Hamiltonian formalism of section \ref{ssec:hamilt} is geared towards canonical quantization, the Lagrangian approach is more convenient for path integrals. Here we show the equivalence between Lagrangian and Hamiltonian formalisms for dynamics on Lie groups at the classical level, at least provided the Hamiltonian has a quadratic kinetic term. The corresponding equivalence at the quantum level is derived in section \ref{sec:G}.

\paragraph{General Hamiltonians.} The first step towards a Lagrangian description of the motion is to define the \i{Hamiltonian action functional}, obtained by integrating the difference between the Liouville one-form \eqref{liouville} and the Hamiltonian function:
\begin{align}
    S[g,p] := \int_0^T \dd t \;\Big( \langle p, \dot g g^{-1}\rangle - H(g,p)\Big).
    \label{e212}
\end{align}
Eqs.~\eqref{3a}--\eqref{3b} extremize this action when the initial positions $g(0)=:g_i$ and $g(T)=:g_f$ are specified.

Momenta can in principle be eliminated from \eqref{e212} by applying the inverse function theorem to eq.~\eqref{3a}, which links momenta and velocities. However, in most applications, the Hamiltonian is the sum of a potential term and a quadratic kinetic term, which renders this elimination straightforward. Assume indeed that $H(g,p)=\frac{1}{2}\langle p, I^{-1}p\rangle+V(g)$, for some potential function $V$ on $G$ and some inertia tensor as in \eqref{euler-arnold}. Then the relation between velocities and momenta is given by eq.~\eqref{velmom}. Plugging this in the Hamiltonian action \eqref{e212} yields the \i{Lagrangian action}
\begin{align}
    S[g]=\int_0^T \dd t \; \Big(\frac{1}{2}\big\langle I(\dot g g^{-1}),\dot g g^{-1}\big\rangle - V(g)\Big),
    \label{laggene}
\end{align}
whose integrand is the usual difference between kinetic and potential energy. Note that the kinetic term is reminiscent of the usual action of nonlinear sigma models on Lie groups (see \i{e.g.}~\cite{Witten:1983ar}). The same Legendre transform will follow from path integration over momenta in section \ref{sec:G}.

\paragraph{Euler-Arnold dynamics as geodesic motion.} For Euler-Arnold systems, the potential term in \eqref{laggene} vanishes, yielding the action functional
\begin{align}
    S[g] =\int_0^T \dd t \; \frac{1}{2}\big\langle I(\dot g g^{-1}),\dot g g^{-1}\big\rangle.
    \label{euler_arnold_action}
\end{align}
Let us derive some on-shell properties of this action to set the stage for path-integral computations. On a classical solution $g_{\text{cl}}(t)$, the Lagrangian in \eqref{euler_arnold_action} reduces to the conserved energy
\begin{equation}
\label{s7}
E_{\text{cl}} 
=
\frac{1}{2}\langle I\dot g_{\text{cl}} g_{\text{cl}}^{-1},\dot g_{\text{cl}} g_{\text{cl}}^{-1}\rangle,
\end{equation}
so the on-shell value of the action is $S_{\text{cl}} = E_{\text{cl}}T$. As in the Hamiltonian approach, the action \eqref{euler_arnold_action} is invariant under right translations $g(t)\mapsto g(t)h$. The associated Noether charge is \eqref{momentmap} with $p = I(\dot g_{\text{cl}} g_{\text{cl}}^{-1})$. 

In fact, the action \eqref{euler_arnold_action} is that of affinely parametrized geodesics for the right-invariant metric $\gamma$ on $G$ given by
\begin{align}
    \gamma_g(u,v) := \langle I \Omega^{\text{R}}_gu, \Omega^{\text{R}}_gv\rangle,
    \label{metric}
\end{align}
where $u,v\in T_gG$ are tangent vectors and $\Omega^{\text{R}}_g=\dd gg^{-1}$ is the right Maurer-Cartan form introduced in eq.~\eqref{liouville}. The length of a path is thus $L[g] = \int_0^T\dd t\,\sqrt{\langle I\dot g g^{-1},\dot g g^{-1}\rangle}$. If the path $g(t)=g_{\text{cl}}(t)$ is on-shell, the integrand here is the square root of the conserved energy \eqref{s7}, so $L[g_{\text{cl}}] = T\sqrt{2E_{\text{cl}}}$. It follows that the on-shell value of the action \eqref{euler_arnold_action} is
\begin{align}
    S_{\text{cl}} = \frac{L[g_{\text{cl}}]^2}{2T}.
    \label{action is length square}
\end{align}
This will govern the leading-order contribution to the quantum propagator, and provides the weights needed for `sums over solitons' in the path integrals of sections \ref{sec:G}--\ref{sec:partition}.

A final note is that all solutions of the Euler-Arnold equations \eqref{velmom}--\eqref{euler_arnold_eom} on the time interval $[0,T]$ can be inferred from solutions on the unit interval $[0,1]$. Indeed, since the path $g_{\text{ref}}(t) := g_{\text{cl}}(Tt)$ defined on $[0,1]$ has the same length and the same image as $g_{\text{cl}}$, it is also a geodesic for the metric \eqref{metric}. Since its energy is constant, it is also affinely parametrized, and therefore solves the Euler-Arnold equations of motion \eqref{velmom}--\eqref{euler_arnold_eom}. It suffices, in this sense, to solve Euler-Arnold equations on a unit time interval, with solutions on $[0,T]$ inferred via the scaling $g_{\text{cl}}(t)=g_{\text{ref}}(t/T)$. Accordingly, the momentum scales as $p_{\text{cl}}(t) = \frac{1}{T}p_{\text{ref}}(t/T)$, as does the Noether charge \eqref{momentmap}, $Q = \frac{1}{T}Q_{\text{ref}}$. This scaling will be used in section \ref{sec:G2}.


\section{Regular representation and Fourier transforms}
\label{sec:review}

This section briefly reviews \cite{Beauvillain:2025ygx}, focussing on the results that are relevant for path integrals. We first recall how an operator algebra stems from the classical Poisson brackets \eqref{e22}, then show how to define its `position' and `momentum' representations. The two are related by noncommutative Fourier transforms---or, more accurately, Fourier coefficients and series when the group $G$ has compact directions. In contrast with \cite{Beauvillain:2025ygx}, stronger emphasis is put on the Dirac notation, which will be extensively used later.

\subsection{Operator algebra for Lie-Poisson systems}
\label{ssec:ops}

The relevant operator algebra is inferred from \eqref{e22} by promoting functions to operators and replacing Poisson brackets by commutators. Its generators are obtained by `putting hats' on coordinate functions of the group---but the latter need not admit global coordinates. To alleviate this complication, one may rely on the exponential map $\exp:\mathfrak{g}\to G:X\mapsto\exp(X)$, assumed to be surjective up to a set of measure zero on $G$. Then define the \i{principal branch of the logarithm} to be the open set in $\mathfrak g$ that contains $0$, on which the exponential is injective with a dense image in $G$, and that is symmetric in the sense that if $X$ belongs to the principal branch, then $-X$ does as well. This defines principal branch coordinates on $G$, which label almost every element of $g\in G$ by a unique element in $X(g)\in\mathfrak{g}$ such that $g=\exp(X(g))$. The identity $e\in G$ sits at the origin $X(e) = 0$, and the inverse is given by $X(g^{-1})=-X(g)$.

Principal branch coordinates, or \i{exponential coordinates} for short, provide an explicit characterization of the operator algebra. Owing to eqs.~\eqref{e22}, the algebra is defined by the commutators
\begin{align}
    [\hat X^i, \hat X^j]= 0,
    \qquad [\hat X^j, \hat p_i] = i \widehat{\mathcal L_iX^j},
    \qquad [\hat p_i, \hat p_j] = -i c^k{}_{ij}\hat p_k.
    \label{commut}
\end{align}
As momentum operators do not commute, one may choose the symmetric ordering to quantize functions of $p$: monomials in $p$ are quantized as sums of all possible orderings of each factor,
\begin{align}
     \mathcal{Q}(p_{i_1}\dots p_{i_N})
     :=
     \frac{1}{N!}\sum_{\sigma\in\text{Sym}(N)} 
     \hat p_{i_{\sigma(1)}}\dots \hat p_{i_{\sigma(N)}}.
     \label{eq: symmetric ordering}
\end{align}
This ensures that exponential functions of momenta are mapped to exponential operators, that is, $\mathcal Q(e^{-i\langle \cdot ,X\rangle}) = e^{-i\langle \hat p, X\rangle}$ for any $X\in\mathfrak{g}$.

Given that the commutators of momenta in \eqref{commut} represent the Lie bracket in $\mathfrak{g}$, the composition of exponential operators reproduces the formula $e^{-i\langle \hat p, X\rangle } e^{-i\langle \hat p,Y\rangle} = e^{-i\langle \hat p, B(X,Y)\rangle}$, where $B(X,Y)$ is the Baker-Campbell-Hausdorff expansion
\begin{align}
B(X,Y)
=
X+Y+\frac{1}{2}[X,Y]+\frac{1}{12}\Big([X,[X,Y]]-[Y,[Y,X]]\Big)+\cdots
\label{baker}
\end{align}
expressed here in terms of Lie algebra brackets (as opposed to commutators). For values of $X$ and $Y$ where the series \eqref{baker} diverges, the global Baker-Campbell-Hausdorff formula is defined by path lifting: take the path $g(t) = \exp(tX)\exp(tY)$ in $G$, and define $Z(t)$ to be the unique continuous path in $\mathfrak{g}$ such that $Z(0)=0$ and $\exp(Z(t))=g(t)$. Then define the \i{global} Baker-Campbell-Hausdorff formula as $B(X,Y):=Z(1)$.

\subsection{Wave functions in position space}
\label{ssec:position}

\paragraph{Wave functions on a Lie group.} The Hilbert space given by the canonical quantization\footnote{Equivalently, geometric quantization with real polarization \cite{woodhouse1997}.} of $G\times \mathfrak{g}^*$ is the space $L^2(G)$ of complex, square-integrable functions on $G$, with elements denoted $\left|\psi\right),\, \left|\phi\right)$, etc. The scalar product is given by integration against the left Haar measure $\dd g_{\text{L}}$,
\begin{align}
    \left(\psi\middle |\phi\right)
    =
    \int_G \dd g_{\text{L}} \;\bar \psi(g)\phi(g) 
    =
    \int\limits_{\substack{\text{principal}\\ \text{branch in $\mathfrak{g}$}}} J_{\text{L}}(X)\dd^n X \; \bar \psi(\exp(X))\phi(\exp(X)),
    \label{posrepscalar}
\end{align}
where $J_{\text{L}}(X)\dd^n X$ is the left Haar measure in exponential coordinates. It is explicitly given by $J_{\text{L}}(X)=\det\Big(\tfrac{1-e^{-\text{ad}_X}}{\text{ad}_X}\Big)$ in terms of the adjoint representation $\text{ad}_X(Y):=[X,Y]$; see eq.~\eqref{e83bis} in appendix \ref{app:B}.

Given $g\in G$, we let $|g)$ be the perfectly localized state at $g$. Its `wave function' maps $h\mapsto\delta_{\text{L}}(g^{-1}h)$, where $\delta_{\text{L}}$ denotes the Dirac distribution at the identity for the left Haar measure. The evaluation of any wave function $|\psi)\in L^2(G)$ can thus be written as $\left(g\middle | \psi\right) = \psi(g)$ upon using the scalar product \eqref{posrepscalar}. Momentum operators generate the left regular representation of $G$ through
\begin{equation}
(g|\hat p_j|\psi) = -i\mathcal L_j\psi(g),
\label{lereg}
\end{equation}
whose exponentiated form is $(g|e^{-i\langle \hat p, X\rangle}|\psi)= \psi(\exp(-X)g)$ for any $X\in\mathfrak{g}$. As for position operators, they act on $L^2(G)$ by multiplication,  as in $(g|\hat V|\psi) = V(g)\psi(g)$. Not all position operators from the algebra \eqref{commut} are allowed to act on $L^2(G)$: since the potential $V$ must be well defined on the group manifold, the corresponding operator should be written as $V(\exp\! \left.(\hat X)\right.)$ in terms of the generators $\hat X$. Another characterization of the subset of operators that are allowed to act on $L^2(G)$ is given below eq.~\eqref{proj}.

\paragraph{Wave functions on a Lie algebra.} It is convenient, for our purposes, to view $L^2(G)$ as a space of periodic functions on $\mathfrak{g}$. To this end, let $L^2(\mathfrak g)$ be the space of square-integrable functions on $\mathfrak{g}$ with the scalar product $\bra{\psi} \ket{\phi}:=\int_\mathfrak g J_{\text{L}}(X)\dd ^n X \;\bar \psi(X)\phi(X)$, where $J_{\text{L}}(X)\dd ^n X$ is the pullback of the left Haar measure, introduced in \eqref{posrepscalar}, by the exponential map. Now define an inclusion $\iota : L^2(G) \hookrightarrow L^2(\mathfrak{g}):|\psi)\mapsto \ket{\iota \psi}$ by
\begin{align}
    \iota\psi(X) := \frac{1}{\sqrt{|\mathbb Z|^r}}\psi(\exp(X)),
    \label{iota}
\end{align}
where $r$ is the the dimension of maximal tori in $G$, which coincides with the rank of $G$ if it is compact. The function $\iota\psi$ is periodic on $\mathfrak{g}$ in the sense that $\iota\psi(X) = \iota\psi(Y)$ if $\exp(X)=\exp(Y)$. Its formal scaling by $1/\sqrt{|\mathbb Z|^r}$ is introduced to ensure that $\iota$ is an (injective) isometry:
\begin{align}
    \bra{\iota\psi}\ket{\iota\phi} 
    &\stackrel{\text{\eqref{posrepscalar}}}{=} 
    \frac{1}{|\mathbb Z|^r}\int_\mathfrak{g}J_{\text{L}}(X)\dd^n X\; \bar{\psi}(\exp(X))\phi(\exp(X))
    \notag
    \\
    &\stackrel{\phantom{\text{\eqref{posrepscalar}}}}{=} 
    \frac{1}{|\mathbb Z|^r}|\mathbb Z|^r\int\limits_{\substack{\text{principal}\\ \text{branch in $\mathfrak{g}$}}}\!\!\!J_{\text{L}}(X)\dd^n X\; \bar{\psi}(\exp(X))\phi(\exp(X)) 
    =
    \left(\psi\middle|\phi\right),
    \label{isomiota}
\end{align}
where the second equality stems from left-invariance of the Haar measure in \eqref{posrepscalar}. This is the sense in which $L^2(G)$ is formally identified with a space of periodic functions on $\mathfrak{g}$. It generalizes the fact that $L^2(\text{U}(1))$ may be viewed as a space of periodic functions on $\mathbb{R}$.

For later reference, it is worth defining perfectly localized states and understanding how they are affected by the inclusion \eqref{iota}. Denote by $|X\rangle$ the function on $\mathfrak{g}$ that maps $Y\mapsto \delta_{\mathfrak{g}}(B(-X,Y))$ in terms of the Baker-Campbell-Hausdorff series \eqref{baker}, with $\delta_{\mathfrak{g}}$ the Dirac distribution at the origin in $\mathfrak{g}$ for the measure $J_{\text{L}}(X)\dd^n X$. One may write $\langle X|\psi\rangle=\psi(X)$ for any wave function $\psi\in L^2(\mathfrak{g})$. Such localized states provide a `basis' of $L^2(\mathfrak{g})$ in the same sense as the usual kets $|x\rangle$ for $L^2(\mathbb{R})$, and the identity operator in $L^2(\mathfrak{g})$ can be written as
\begin{align}
    \operatorname{id}
    =
    \int_\mathfrak{g}J_{\text{L}}(X)\dd^nX \; \ket{X}\bra{X},
    \label{idpos}
\end{align}
with $J_{\text{L}}(X)\dd^n X$ the (exponential pullback of the) left Haar measure as in \eqref{posrepscalar}. This will be crucial for path integrals.

Now consider the effect of the inclusion \eqref{iota} on perfectly localized states. For almost every $g\in G$, the inclusion of $\left|g\right)$ into $L^2(\mathfrak g)$ generalizes the Dirac comb distribution,
\begin{align}
    \ket{\iota g} 
    =
    \frac{1}{\sqrt{|\mathbb Z|^r}}\sum_{X\in \operatorname{Logs}(g)} \ket{X} 
    =
    \frac{1}{\sqrt{|\mathbb Z|^r}}\sum_{n^1,\dots, n^r\in\mathbb{Z}} \ket{X + 2\pi n^ia_i(X)}.
    \label{inclusion loc}
\end{align}
Here $\operatorname{Logs(g)}:=\{X\in \mathfrak{g}\,|\,\exp(X)=g\}$ denotes the set of logarithms of $g$ in $\mathfrak{g}$. For a generic element in $G$, this set can be characterized as follows. Let $T\cong(\mathbb{S}^1)^r$ be the maximal torus passing through $g$. The Lie algebra $\mathfrak{t}$ of $T$ contains the corresponding element $X$ in the principal branch, so let $\{a_i(X)|i=1,\dots,r\}$ be a basis of $\mathfrak{t}$ such that $X\propto a_1(X)$, with $\exp(2\pi a_i(X))=e$ but $\exp(\theta a_i(X))\neq e$ for all $\theta \in (0,2\pi)$. Since $g$ is generic, its logarithms are discrete translations along the Lie algebra of $T$:
\begin{align}
    \operatorname{Logs}(g)
    =
    \{X+2\pi n^i a_i(X)\,|\, n^i\in \mathbb{Z}\}
    \qquad\text{for almost any $g\in G$}.
    \label{genlogs}
\end{align}
This justifies \i{a posteriori} the scaling \eqref{iota} of periodic functions by $1/\sqrt{|\mathbb Z|^r}$, as well as the simplification in \eqref{isomiota}, since the branches of the logarithm are labelled by elements of $\mathbb Z^r$.

Note that one can also go the other way by mapping a (nonperiodic) function in $L^2(\mathfrak{g})$ on a function in $L^2(G)$. This is analogous to the way that functions on the real line are made periodic by summing over the images of their discrete translations. To that end, define the projector\footnote{\label{foo}Note that the identity $e\in G$ is \i{not} a `generic' element of $G$; its set of logarithms is much larger than \eqref{genlogs}. One should therefore be careful in defining the normalization $|\mathcal I|=|\operatorname{Logs}(e)|$ in eq.~\eqref{proj}. We briefly return to this point in section \ref{sec:partition}.}
\begin{align}
    \hat P_{\mathcal I} 
    :=
    \frac{1}{|\mathcal I|}\sum_{Y\in \mathcal I} e^{-i\langle \hat p, Y\rangle},
    \qquad \text{with}\qquad
    \mathcal I
    :=
    \operatorname{Logs}(e),
    \label{proj}
\end{align}
which provides the desired map $L^2(\mathfrak{g})\to L^2(G)$, in the sense that $\iota(L^2(G)) = \sqrt{|\mathbb Z|^r} \hat P_\mathcal I\cdot L^2(\mathfrak{g})$ in terms of the inclusion \eqref{iota}. It follows that an operator $\hat O$ acting on $L^2(G)$ is one that stabilizes $\sqrt{|\mathbb Z|^r} \hat P_\mathcal I\cdot L^2(\mathfrak{g})$ when acting on $L^2(\mathfrak g)$. In addition, any such operator commutes with translations by logarithms of the identity,
\begin{align}
    \left[\hat O, e^{-i\langle \hat p, Z\rangle}\right]=0 \qquad \forall~Z\in\operatorname{Logs}(e).
    \label{logsidcommute}
\end{align}
Here translations in $L^2(\mathfrak{g})$ are given by the Lie algebraic version of \eqref{lereg}, namely
\begin{align}
    \bra{X}e^{-i\langle \hat p, Y\rangle}\ket{\psi} = \psi(B(-Y,X))
    \label{translation}
\end{align}
in terms of the Baker-Campbell-Hausdorff expansion \eqref{baker}. Eq.~\eqref{logsidcommute} is the Lie-group analogue of the (trivial) fact that $2\pi$-periodic Hamiltonians on $L^2(\mathbb R)$ are those that commute with translations by integer multiples of $2\pi$.

\subsection{Momentum space and Fourier transforms}
\label{ssec:mom}

\paragraph{Fourier transforms.} The Hilbert space $L^2_\star(\mathfrak g^*)$ of the momentum representation of the algebra \eqref{commut} consists of functions of $p\in \mathfrak g^*$. Its elements are denoted $\Phi$, $\Psi$, etc.~and the scalar product reads $\bra{\Psi}\ket{\Phi} := \int_{\mathfrak{g}^*}\frac{\dd^np}{(2\pi)^n}\;\bar \Psi(p)\star \Phi(p)$, where $\star$ is the \i{Gutt star product} defined by the relation \cite{Gutt,Dito,Kathotia}
\begin{align}
    e^{-i\langle p, X\rangle}\star e^{-i\langle p, Y\rangle} 
    =
    e^{-i\langle p, B(X,Y)\rangle},
    \label{guttdef}
\end{align}
with $B(X,Y)$ the (global) Baker-Campbell-Hausdorff formula \eqref{baker}. On this Hilbert space, position operators act by differentiation $(\hat X^j\Psi)(p) = i\partial_{p_j}\Psi$, while momenta act by star multiplication: if $K$ is a function of $p$ and $\hat K = \mathcal Q(K)$ is its symmetric quantization \eqref{eq: symmetric ordering}, then $(\hat K\Psi)(p):= K(p)\star \Psi(p)$.

The noncommutative Fourier transform \cite{Beauvillain:2025ygx,Oriti:2011ac} provides a bijective isometry $F:L^2(\mathfrak g)\to L^2_\star(\mathfrak g^*)$ that commutes with the action of operators, meaning that $F[\hat O\psi] = \hat O F[\psi]$. It is defined as
\begin{align}
    F[\psi](p):= \int_\mathfrak gJ_{\text{L}}(X)\dd^n X\; e^{-i\langle p, X\rangle} \psi(X),
    \label{defFT}
\end{align}
where $J_{\text{L}}(X)\dd^n X$ is again the left Haar measure in exponential coordinates, as in eq.~\eqref{posrepscalar}. The inverse Fourier transform is
\begin{align}
    F^\dagger[\Psi](X) = \int_{\mathfrak g^*}\frac{\dd^np}{(2\pi)^n}e^{i\langle p, X\rangle}\star \Psi(p).
    \label{FTinv}
\end{align}
Thus, the Hilbert spaces $L^2(\mathfrak g)\cong L^2_\star(\mathfrak g^*)$ may safely be identified.

It is essential, for path integrals, to have both position-space and momentum-space representations of the identity operator in $L^2(\mathfrak{g})\cong L^2_{\star}(\mathfrak{g}^*)$. In position space, the identity is given by \eqref{idpos} in terms of the localized states $|X\rangle$. The analogous formula in momentum space is obtained by defining the (non-normalizable) state $\ket{p}\in L^2(\mathfrak{g})$ by $\bra X\ket{p} = e^{i\langle p, X\rangle}$. The identity \eqref{idpos} and the Fourier transform \eqref{defFT} then yield $\bra{p}\ket{\psi}=F[\psi](p)$, so $\ket{p}$ is a perfectly localized state in momentum space. In fact, it is the `wave function' that maps $q\mapsto\delta_\star(q,p)$, where $\delta_\star$ is the distribution on $\mathfrak g^*$ such that
\begin{align}
    \int_{\mathfrak{g^*}} \frac{\dd^np}{(2\pi)^n}\;\delta_\star(p,q)\star_q \Psi(q) 
    =
    \Psi(p)
    \label{deltastar}
\end{align}
for any momentum-space wave function $\Psi$. It finally follows from eq.~\eqref{FTinv} that
\begin{align}
    \operatorname{id} = \int_{\mathfrak g^*}\frac{\dd^n p}{(2\pi)^n}\ket{p}\star \bra{p},
    \label{idmom}
\end{align}
which is the sought-for momentum-space representation of the identity. Again, this will be essential for path integrals.

\paragraph{Fourier coefficients and series.} Fourier coefficients on $L^2(G)$ are obtained by acting on both sides of the Fourier transform \eqref{defFT} with the projector \eqref{proj}. As mentioned there, acting with $\hat P_\mathcal I$ on $L^2(\mathfrak g)$ yields, after rescaling, a Hilbert space isomorphic to $L^2(G)$. One can similarly define the projected momentum Hilbert space as $\sqrt{|\mathbb Z|^r}\hat P_\mathcal I \cdot L^2_\star(\mathfrak{g}^*)$.\footnote{In \cite{Beauvillain:2025ygx}, the scaling by $\sqrt{|\mathbb Z|^r}$ was kept implicit in the definition of the `small' momentum representation. Since the exposition here is much more concise, we keep the scaling explicit for clarity.} The end result is that noncommutative Fourier coefficients $F_\mathcal I: L^2(G)\rightarrow \sqrt{|\mathbb Z|^r}\hat P_\mathcal I \cdot L^2_\star(\mathfrak{g}^*)$ are given by
\begin{align}
    F_\mathcal I[\psi](p) 
    :=
    \sqrt{|\mathbb Z|^r}
    \hat P_\mathcal IF[\psi](p)
    =
    \sqrt{|\mathbb Z|^r}
    \int\limits_{\substack{\text{principal}\\ \text{branch in $\mathfrak{g}$}}}\!\!\! J_{\text{L}}(X)\dd^nX \; E_\mathcal I(X,p)\psi(\exp(X)),
\end{align}
where we let
\begin{align}
    E_\mathcal I(X,p) =e^{-i\langle p, X\rangle}\sum_{k_1,\dots,k_r\in \mathbb{Z}} \prod_{i=1}^r\frac{\delta(\langle p, a_i(X)\rangle-k_i)}{|\mathbb Z|}
    \label{coeffs}
\end{align}
in terms of the maximal torus directions $a_i(X)$ defined below \eqref{genlogs}. Similarly, Fourier series are given by $F_{\mathcal{I}}^{\dagger}[\Phi](g)
      =
      \sqrt{|\mathbb Z|^r}\int_{\mathfrak{g}^*} 
      \frac{\dd^n p}{(2\pi)^n}\;\overline{E_\mathcal{I}(g,p)}\star\Phi(p)$.


\section{Path integrals on Lie groups}
\label{sec:G}

This section proves the equivalence between canonical quantization and path integral quantization for Lie groups, by expressing the quantum propagator in $L^2(G)$ as a path integral. We rely on the $G=\text{U}(1)$ case, detailed in appendix \ref{app:U1}, as inspiration. Thus, we first use the inclusion \eqref{iota} to rewrite sums over paths in $G$ as `decompactified' sums over paths in $\mathfrak{g}$ with additional `winding numbers'. We then exploit noncommutative Fourier transforms to repeatedly insert the identity \eqref{idpos}--\eqref{idmom} inside the decompactified propagators. The result is eq.~\eqref{e316} for the propagator, which we eventually rewrite in terms of the right (rather than left) measure and relate to the Lagrangian approach of section \ref{sselag}.

\subsection{Building the path integral}
\label{ssec:trotter}

\paragraph{From paths in $G$ to paths in $\mathfrak{g}$.} Given a Hamiltonian operator $\hat H$ acting on $L^2(G)$, our goal is to express the propagator (or time evolution kernel) $(g_f | e^{-i\hat H T}| g_i)$ as a path integral. The first step is to decompactify the space on which the propagator is defined, as is commonly done in the U(1) case (see appendix \ref{app:U1}). This is straightforward with our current formalism, as the inclusion \eqref{iota} can be used to write
\begin{align}
       (g_f | e^{-i\hat H T}| g_i)
       =
       \frac{1}{|\mathbb{Z}|^r}\sum_{Y_f\in \operatorname{Logs}(g_f)}\sum_{Y_i\in \operatorname{Logs}(g_i)} \bra{Y_f}e^{-i\hat H T}\ket{Y_i}.
       \label{42}
\end{align}
Since $Y_i$ here is a logarithm of $g_i$, it can be obtained by translating the principal branch logarithm $X_i$ of $g_i$. There exists, in other words, a $Z(Y_i)\in \operatorname{Logs}(e)$ such that $Y_i = B(Z(Y_i),X_i)$, which by \eqref{translation} means that $\ket{Y_i} =e^{i\langle \hat p, Z(Y_i)\rangle}\ket{X_i}$. But the Hamiltonian $\hat H$ has a well-defined action on $L^2(G)$, so eq.~\eqref{logsidcommute} ensures that it commutes with translations by logarithms of the identity, and eq.~\eqref{42} can be recast as \begin{equation}
(g_f | e^{-i\hat H T}| g_i) 
=
\sum_{Y_f\in \operatorname{Logs}(g_f)}\bra{Y_f}e^{-i\hat H T}\ket{X_i},
\label{e312}
\end{equation}
where $g_i=\exp(X_i)$ and $X_i$ belongs to the principal branch. Note that the formal $\frac{1}{|\mathbb{Z}|^r}$ prefactor simplifies, as each $\ket{Y_i}$ from \eqref{42} yields the same contribution, of which there are $|\mathbb Z|^r$. The key virtue of \eqref{e312} is to express the propagator on $L^2(G)$ as a sum of decompactified kernels on $L^2(\mathfrak g)$. It remains to express each decompactified propagator $\bra{Y_f}e^{-i\hat H T}\ket{X_i}$ as a path integral. 

\paragraph{Time discretization.} The next step in the path integral derivation is to cut the time interval $[0,T]$ into a large number $N$ of smaller segments of size $\Delta T=T/N$, and insert resolutions of the identity between neighbouring segments. This is readily achieved thanks to noncommutative Fourier transforms, with the identity expressed either in position space \eqref{idpos} or in momentum space \eqref{idmom}. Letting $X_0 = X_i$ and $X_N = Y_f$, the result is
\begin{align}
       \bra{Y_f}e^{-i\hat H T}\ket{X_i} 
       =
       \int \prod_{j=1}^{N-1}\left(\tfrac{J_{\text{L}}(X_j)\dd^n X_j\dd^n p_j}{(2\pi)^n}\right)\tfrac{\dd^n p_0}{(2\pi)^n} 
       \prod_{k=0}^{N-1} \underbrace{\bra{X_{k+1}}\ket{p_k}\star_{p_k}\!\!\bra{p_k}e^{-i\hat H T/N}\ket{X_k}}_{\displaystyle:=F_k},
       \label{e38}
\end{align}
where we stress that each integral over $X_j$ runs over the whole Lie algebra $\mathfrak{g}$, and that the one over $p_j$ runs over the whole dual $\mathfrak{g}^*$.

There are two major differences between eq.~\eqref{e38} and its standard `Abelian' analogue in $L^2(\mathbb{R})$ or $L^2(\text{U}(1))$. The first is the Jacobian $J_{\text{L}}(X_j)$ that affects each time step $j$, which is unavoidable with path integrals on curved spaces. The second is the star product entering each factor $F_k$, whose effects we now study.

Expanding the exponential in \eqref{e38} at large $N$ yields
\begin{equation}
       F_k
       \sim
       \bra{X_{k+1}}\ket{p_k}\star \bra{p_k}\Big(\!1-\tfrac{i\hat H T}{N}\!\Big)\ket{X_k}
       =
       \bra{X_{k+1}}\ket{p_k}\star\bra{p_k}\ket{X_k}\star\Big(\!1-\tfrac{iT}{N}\!\bra{X_k}\ket{p_k}\star \bra{p_k}\hat H\ket{X_k}\!\Big),
       \label{e40}
\end{equation}
where we used $\bra{p}\ket{X}\star \bra{X}\ket{p} = e^{-i\langle p, X\rangle}\star e^{i\langle p, X\rangle} = 1$. The symbol of the Hamiltonian is then defined as 
\begin{align}
    H(X,p):= \bra{X}\ket{p}\star \bra{p}\hat H\ket{X},
    \label{symbol}
\end{align}
which is generally affected by the star product. Plugging this into eq.~\eqref{e40} and using the Gutt star product \eqref{guttdef} yields the large $N$ expression
\begin{align}
       F_k\sim e^{i\langle p_k, B(-X_k, X_{k+1})\rangle}\star\Big(1-\tfrac{iT}{N}H(X_k,p_k)\Big).
       \label{4.8}
\end{align} 
Before taking the $N\rightarrow \infty$ limit, rewrite $X_{k+1} = X_k + \frac{T}{N}\dot X_k$, with the discrete derivative defined as $\dot X_k := \frac{X_{k+1}-X_k}{\Delta T}$. This converges to the derivative $\dot X(t)$ in the $N\rightarrow\infty$ limit, provided $X_k$ interpolates a differentiable path $X(t)$. As a result, one can relate the Baker-Campbell-Hausdorff formula to the left Maurer-Cartan form $\Omega^{\text{L}}_g := g ^{-1}\dd g$ via $B(-X,X + \epsilon \dot X) = \epsilon(\exp^*\Omega^{\text{L}})_X(\dot X) + O(\epsilon^2)$, where $\exp^*$ denotes the pullback of a form from $G$ to $\mathfrak{g}$ by the exponential map. We lighten the notation by writing $(\exp^*\Omega^{\text{L}})_X=:\Omega^{\text{L}}_X$ (and similarly for $\Omega^{\text{R}}$), being understood that the subscript indicates whether the Maurer-Cartan form lives on $G$ (as in $\Omega_g$) or on $\mathfrak{g}$ (as in $\Omega_X$). The factor \eqref{4.8} thus becomes
 \begin{align}
       F_k 
       \sim 
       e^{\frac{iT}{N}\langle p_k, \Omega^{\text{L}}_{X_k}(\dot X_k)\rangle}\star \Big(1-\frac{iT}{N}H(X_k,p_k)\Big)
       \sim
       e^{\frac{iT}{N}\left(\langle p_k, \Omega^{\text{L}}_{X_k}(\dot X_k)\rangle - H(X_k,p_k)\right)}
       \label{e312bis}
 \end{align}
at leading order in $1/N$, upon expanding the exponential and noting that star products only contribute subleading terms. Plugging this into the discretized propagator \eqref{e38} leads to
 \begin{equation}
 \begin{split}
       &\bra{Y_f}e^{-i\hat H T}\ket{X_i} 
       =\\
       &\int \prod_{j=1}^{N-1}\left(\frac{J_{\text{L}}(X_k)\dd^n X_k\dd^n p_k}{(2\pi)^n}\right) \frac{\dd^n p_0}{(2\pi)^n} 
       \exp\bigg[\frac{iT}{N}\sum_{k=0}^{N-1} \Big(\big\langle p_k,\Omega^{\text{L}}_{X_k}(\dot X_k)\big\rangle - H(X_k,p_k)\Big)\bigg].
       \end{split}
       \label{e44}
 \end{equation}

\paragraph{Continuum limit and left action functional.} In the limit $N\rightarrow\infty$, the Riemann sum in the exponential of \eqref{e44} converges to the action functional
 \begin{align}
       S[X,p] 
       = \int_0^T \dd t\; \Big(\big\langle p, \Omega^{\text{L}}_X(\dot X)\big\rangle - H(X,p)\Big)
       = \int_0^T \dd t\; \Big(\langle p, g^{-1}\dot g\rangle - H(g,p)\Big),
       \label{e44b}
 \end{align}
where $g=\exp(X)$. This is \i{not} the action \eqref{e212} since left and right Maurer-Cartan forms generally differ; we resolve this apparent discrepancy in section \ref{ssec:symb}. Now introduce the formal path integral measure
\begin{equation}
\mathcal{D}X_{\text{L}}\mathcal Dp 
:=
\lim_{N\rightarrow\infty}\frac{\dd^n p_0}{(2\pi)^n}\prod_{j=1}^{N-1}\frac{J_{\text{L}}(X_k)\dd^n X_k\dd^n p_k}{(2\pi)^n}
\label{pamela}
\end{equation}
to rewrite the propagator \eqref{e44} as a continuous-time path integral. Also including the sum over logarithms in the full propagator \eqref{e312}, we find
\begin{align}
(g_f| e^{-i\hat H T}|g_i)
=
\sum_{Y_f\in \operatorname{Logs}(g_f)} \int_{X(0)= X_i}^{X(T)=Y_f}\mathcal{D}X_{\text{L}}\mathcal Dp\;e^{iS[X,p]},
\label{e316}
\end{align}
where $S$ is the action \eqref{e44b}. This is our final path-integral expression for the propagator in $L^2(G)$, for any Hamiltonian operator $\hat H$, whose symbol is defined in \eqref{symbol}. We stress again that the integrals over $X$ and $p$ in \eqref{e316} are `decompactified' in the sense that they respectively run over the entire Lie algebra $\mathfrak{g}$ and its dual $\mathfrak{g}^*$.

Eq.~\eqref{e316} for the time-evolution kernel has the same structure as the usual path integral for a particle on a circle (see appendix \ref{app:U1}), with the sum over winding numbers replaced by a sum over logarithms of the final group element. This last sum can also be interpreted as a sum over windings, here along directions in a maximal torus. There are, however, two key differences between eq.~\eqref{e316} and standard U(1) path integrals---we anticipated them above eq.~\eqref{e40}. First, the path integral measure \eqref{pamela} is not flat and involves a Jacobian $J_{\text{L}}(X(t))$ at each time step. This affects two-loop calculations in section \ref{ssecTwoLoop}. Second, the symbol \eqref{symbol} of the Hamiltonian involves a star product, which entails counter-intuitive effects. Let us address this point right away.

\subsection{From left to right Haar measures}
\label{ssec:symb}

The symbol of the Hamiltonian \eqref{symbol} involves a star product; evaluating it is therefore nontrivial. For definiteness, consider a quantum Hamiltonian $\hat H = \hat K(p) + \hat V(g)$, which is the sum of a kinetic term and a potential term. The generalization to arbitrary Hamiltonians $H(X,p)$ is straightforward because the $X$ operators commute in \eqref{commut}.

\paragraph{Symbols: potential and kinetic.} The symbol of the potential term is just $V(X,p) = \bra{X}\ket{p}\star\bra{p}\hat V\ket{X} = \bra{X}\ket{p}\star\bra{p}\ket{X}V(X) = V(X)$. For the kinetic term, on the other hand, recall from the definition below eq.~\eqref{guttdef} that any kinetic operator acts on wave functions through star multiplication $(\hat K \Psi)(p) = K(p)\star \Psi(p)$ in momentum space. Here $K(p)$, with one argument, denotes the classical function of $p$ whose symmetric quantization \eqref{eq: symmetric ordering} gives the operator $\hat K$. It must not be confused with the function $K(X,p)$, with two arguments, which denotes the symbol \eqref{symbol} of the operator $\hat K$. These two functions are related as follows. Using \eqref{deltastar}--\eqref{idmom}, note first that
\begin{align}
       \bra{X}\hat K\ket{p} &= \int \frac{\dd^n q}{(2\pi)^n} \bra{X}\ket{q}\star_q\bra{q}\hat K\ket{p} 
       \notag
       \\
       &= \int \frac{\dd^n q}{(2\pi)^n}\; e^{i\langle q, X\rangle}\star_q K(q)\star_q \delta_\star(q,p) = e^{i\langle p,X\rangle}\star K(p),
\end{align}
so that the symbol of the kinetic term reads
\begin{equation}
\label{symko}
K(X,p) 
=
\bra{X}\ket{p}\star \overline{\bra{X}\hat K \ket{p}} 
=
e^{i\langle p,X\rangle}\star K(p) \star e^{-i\langle p,X\rangle}.
\end{equation}
This is not yet explicit, but it can be simplified using the linearity of the star product \eqref{guttdef} and the fact that plane waves form a basis of the space of functions of $p$. Explicitly, replacing $K(p)$ on the far right-hand side of \eqref{symko} by an exponential, one has
\begin{align}
       e^{i\langle p,X\rangle}\star e^{-i\langle p,Y\rangle} \star e^{-i\langle p,X\rangle} = e^{-i\langle p, B(-X,B(Y,X))\rangle} = e^{-i\langle p, \text{Ad}_{\exp(-X)}Y\rangle}=e^{-i\langle \text{Ad}^*_{\exp(X)}p,Y\rangle}.
\end{align}
This implies that the symbol \eqref{symko} finally becomes
\begin{align}
       K(X,p) = K(\text{Ad}^*_{\exp(X)}p).
       \label{e50}
\end{align}
Note that this mixes momenta and positions, even though the operator $\hat K$ only depends on momenta!

\paragraph{From left to right measures; modular function.} The subtlety in the symbol \eqref{e50} is closely related to the difference between the action functionals \eqref{e44b} and \eqref{e212}. The manifest separation between kinetic and potential energies can be recovered through a change of variables in the path integral: letting $\pi(t) := \text{Ad}^*_{\exp(X(t))}p(t)$ in eq.~\eqref{e316} yields the propagator
\begin{equation}
\begin{split}
       &(g_f| e^{-i\hat H T}|g_i)
       =\\
       &\frac{1}{\Delta(g_i)}\sum_{Y_f\in \operatorname{Logs}(g_f)} \int\limits_{X(0)= X_i}^{X(T)=Y_f}\mathcal{D}X_{\text{R}}\mathcal D\pi \; \exp\bigg[i\int_0^T\dd t \Big(\langle \pi, \Omega^{\text{R}}_X(\dot X)\rangle - K(\pi)-V(X)\Big)\bigg].
       \end{split}
       \label{e52b}
\end{equation}
Everything is now written in terms of the \i{right} Haar measure rather than the left one: $J_{\text{R}}(X)\dd^nX$ is the right Haar measure in exponential coordinates (see eq.~\eqref{e83} in appendix \ref{app:B}), the path integral measure is defined as in \eqref{pamela} with $J_{\text{L}}$ replaced by $J_{\text{R}}$, $\Omega^{\text{R}}_g:=\dd gg^{-1}$ is the right Maurer-Cartan form as in \eqref{liouville}, and
\begin{equation}
\label{mofo}
\Delta(g):= \frac{J_{\text{L}}(X)}{J_{\text{R}}(X)}=\frac{\dd g_{\text{L}}}{\dd g_{\text{R}}}
\end{equation}
is the modular function on $G$ (see eq.~\eqref{modular function} in appendix \ref{app:B}), \i{i.e.} the Radon-Nikodym derivative of the left Haar measure with respect to the right one. To obtain \eqref{e52b}, we used the relation $\Omega^{\text{R}}_g = \dd g \,g^{-1} = g(g^{-1}\dd g) g^{-1}=\text{Ad}_{g}\Omega^{\text{L}}_g$ between left and right Maurer-Cartan forms, as well as the property $\det(\text{Ad}_{\exp(X)})J_{\text{L}}(X) = J_{\text{R}}(X)$. The action that now appears in the path integral is the same as in eq.~\eqref{e212}, with a classical Hamiltonian $H(g,p)=K(p)+V(g)$.

Note that the extra $1/\Delta(g_i)$ factor in eq.~\eqref{e52b} can be understood by going back to the discretized path integral \eqref{e44}. The latter contains one more integration in momentum space than there are in position space: the one over $p_0$. As a result, the change of variable $\pi_0 = \text{Ad}^*_{\exp(X_i)}p_0$ produces a Jacobian $\det(\text{Ad}_{\exp(X_i)}) = 1/\Delta(g_i)$ that is not reabsorbed in a change from left- to right-invariant Haar measures. 

\subsection{Integrating out momenta}

The action functional in the path integral \eqref{e52b} is Hamiltonian, in that it depends on both phase space variables $X$ and $p$. In order to rewrite the path integral in terms of the Lagrangian action \eqref{laggene}, one needs to integrate out momenta. We do this by further assuming that the kinetic part of the Hamiltonian is quadratic. More precisely, assume that the kinetic term reads $\hat K = \frac{1}{2}\langle \hat p , I^{-1} \hat p\rangle$ in terms of some inertia tensor $I$, as in eq.~\eqref{euler-arnold}. The latter is symmetric, so $\hat K$ is the symmetric quantization of the phase space function $K(p) = \frac{1}{2}\langle p , I^{-1} p\rangle$ and eq.~\eqref{e52b} yields
\begin{equation}
\begin{split}
        &(g_f| e^{-i\hat H T}|g_i)
        =\\
        &\frac{1}{\Delta(g_i)}\sum_{Y_f\in \operatorname{Logs}(g_f)}
        \int\limits_{X(0)= X_i}^{X(T)=Y_f}
        \mathcal{D}X_{\text{R}}\mathcal D\pi 
        \;
       \exp\bigg[i\int_0^T\dd t\Big( \big\langle \pi, \Omega^{\text{R}}_X(\dot X)\big\rangle - \frac{1}{2}\langle \pi, I^{-1}\pi\rangle -V(X)\Big)\bigg].
       \end{split}
\end{equation}
Gaussian integration over $\pi$ can now be performed, leading to
\begin{equation}
\begin{split}
        &(g_f| e^{-i\hat H T}|g_i) = \\
        &\frac{\mathcal N}{\Delta(g_i)}\sum_{Y_f\in \operatorname{Logs}(g_f)}
        \int\limits_{X(0)= X_i}^{X(T)=Y_f}
        \mathcal{D}X_{\text{R}}
        \;
        \exp\bigg[i\int_0^T\dd t \Big(\frac{1}{2}\big\langle I (\Omega^{\text{R}}_X(\dot X)), \Omega^{\text{R}}_X(\dot X)\big\rangle-V(X)\Big)\bigg].
        \end{split}
        \label{e55}
\end{equation}
Here the divergent normalization $\mathcal N$ formally satisfies
\begin{equation}
\mathcal N\propto\prod_{0 \leq t < T} \sqrt{\det I},
\label{s16}
\end{equation}
and the remaining path integral measure is
\begin{equation}
\label{pamelag}
\mathcal{D}X_{\text{R}}=\lim_{N\rightarrow\infty}\prod_{j=1}^{N-1}J_{\text{R}}(X_k)\dd^n X_k,
\end{equation}
as follows from the phase space measure \eqref{pamela}. The action functional in \eqref{e55} is precisely the Lagrangian one \eqref{laggene}, written in exponential coordinates.

Eq.~\eqref{e55} for the time-evolution kernel will be our starting point for saddle point approximations around classical trajectories. In the literature (see \i{e.g.}~\cite{Raasakka:2011np} and references therein), path integrals on Lie groups are more commonly written as
\begin{equation}
\int_{g(0)=g_i}^{g(T)=g_f}\mathcal{D}g_{\text{R}}\;
\exp\bigg[i\int_0^T \dd t\Big(\frac{1}{2}\langle I \dot gg ^{-1}, \dot gg ^{-1}\rangle-V(g)\Big)\bigg],
\end{equation}
where $\mathcal D g_{\text{R}}$ is the formal product of infinitely many right Haar measures, one for each time $t\in[0,T]$. This is indeed very close to our eq.~\eqref{e55}, but we refrain from using such a simpler expression for two reasons. First, it does not explain why integration over the perturbations around classical trajectories should be noncompact, as is crucially the case when writing the path integral in U(1) as a path integral in $\mathbb{R}$ with extra winding numbers (see appendix \ref{app:U1}). Second, it does not provide a clear counting of the classical trajectories to be summed over. Our eq.~\eqref{e55}, by contrast, clearly separates global/topological information, encoded in the sum over logarithms, from the local degrees of freedom running in decompactified path integrals.


\section{Semiclassical Euler-Arnold systems}
\label{sec:G2}

We now focus on the perturbative path integral computation of the propagator of quantum Euler-Arnold systems---the most relevant examples of Lie-Poisson dynamics \cite{Khesin,Gripaios:2015pfa}. The corresponding classical equations of motion \eqref{velmom}--\eqref{euler_arnold_eom} describe geodesics in $G$, so the propagator boils down to a heat kernel. This makes it possible to compare our framework to known results on heat kernel expansions on Lie groups \cite{Mori:2019,Craddock:2017,David:2010,avramidi:2006,Vassilevich:2003xt}.

The section is organized as follows. We first show, for purely kinetic Hamiltonians, that the propagator from the identity determines all other propagators by virtue of right-invariance. We then initiate perturbation theory by expanding the path integral around classical trajectories, going up to (and including) two-loop order.

\subsection{Propagators from the identity}
\label{ssec:rightinv}

Lie-Poisson Hamiltonians are invariant under right translations (recall the conserved charge \eqref{momentmap}). This remains true at the quantum level. Indeed, any Hamiltonian $\hat H(p)$ trivially commutes with right translations since it is wholly written in terms of $\hat p$, which acts on $L^2(G)$ by infinitesimal \i{left} translations \eqref{lereg}. Translation invariance therefore suggests that the propagator at, say, $g_i=e$ contains all the information about time evolution.

This is indeed the case, but the proof involves a subtlety having to do with left and right Haar measures. Indeed, define for any $h\in G$ the operator $(\hat R_h\psi)(g):=\psi(gh)$ that acts on wave functions on $G$ by right translations. Its action on a perfectly localized state can be found by computing the scalar product $(\psi|\hat R_l|{h}) = \int \dd g_{\text{L}} \; \bar \psi(g) \delta_{\text{L}}(h^{-1}gl)=\int \dd u_{\text{R}}\; \Delta(u)\bar \psi(hu)\delta_{\text{L}}(ul)$, where we used eq.~\eqref{posrepscalar}, changed variables to $u=h^{-1}g$ and related the left and right Haar measures through the modular function \eqref{mofo}. Changing variable once more to $v = ul$ and using the fact that the modular function is multiplicative to come back to left Haar measure, one gets
\begin{align}
(\psi|\hat R_l|{h}) 
=
\int \dd v_{\text{L}}\; \Delta(l^{-1})\bar\psi(hvl^{-1})\delta_{\text{L}}(v) 
=
\bar{\psi}(hl^{-1})\Delta(l^{-1}) 
=
({\psi}|\Delta(l^{-1})|{hl^{-1}}),
    \label{328}
\end{align}
so that $\hat R_h\left|g\right) = \Delta(h^{-1})\left|gh^{-1}\right)$. The adjoint of right translations can be computed in the same way, and one finds 
$\hat R_h^\dagger = \Delta(h^{-1})\hat R_{h^{-1}}$. This can be used to express any kernel \eqref{42} of Lie-Poisson systems in terms of one that starts at the identity. Indeed, by eq.~\eqref{328}, one can rewrite $\left|g_i\right)=\Delta(g_i^{-1})\hat R_{g_i^{-1}}\left|e\right)$. Since any Lie-Poisson Hamiltonian commutes with right translations, the propagator reads
\begin{align}
    (g_f| e^{-i\hat H T}|g_i)  
    = 
    \Delta(g_i^{-1}) 
    (g_fg_i^{-1}| e^{-i\hat H T}|e).
    \label{e right invariance}
\end{align}
We therefore restrict attention, without loss of generality, to the propagator \eqref{e55} starting at the identity:
\begin{align}
    (g_f|e^{-i\hat H T}|e)
    = 
        \mathcal N\sum_{Y_f\in \operatorname{Logs}(g_f)}
        \int_{X(0)= 0}^{X(T)=Y_f}
        \mathcal{D}X_{\text{R}}
        \;
        e^{i\int_0^T\dd t \frac{1}{2}\langle I (\Omega^{\text{R}}_X(\dot X)),\Omega^{\text{R}}_X(\dot X)\rangle},
        \label{euea}
\end{align}
where we used $\Delta(e)=1$. Note that eq.~\eqref{e right invariance} is consistent with the dependence of the propagator \eqref{e55} on the modular function.

\subsection{Setting up perturbation theory}
\label{ssec:settingup}

We wish to perform a saddle point approximation around classical trajectories in the propagator \eqref{euea}, so as to write it as an asymptotic series in $\hbar T(=T)$. To set this up, we now expand the action \eqref{euler_arnold_action} around classical trajectories and account for the curved measure by introducing Faddeev-Popov ghosts.

\paragraph{Perturbed action functional.} Since the action in \eqref{euea} is that of eq.~\eqref{euler_arnold_action}, it has a saddle point at the equations of motion \eqref{velmom}--\eqref{euler_arnold_eom}. Recall that many Euler-Arnold systems are classically integrable \cite{Khesin}, providing a convenient starting point for semiclassical analysis. Let us therefore assume that all solutions $g_{\text{cl}}(t)$ of \eqref{velmom}--\eqref{euler_arnold_eom}, linking $e$ to $g_f$ in time $T$, are known. Denote by $X_{\text{cl}}$ the lifting of a given classical path $g_{\text{cl}}$ to the Lie algebra, \i{i.e.}~the unique continuous path in $\mathfrak{g}$ such that $\exp(X_{\text{cl}}(t))=g_{\text{cl}}(t)$ and $X_{\text{cl}}(0)=0$ (since $g_{\text{cl}}(0)=e$).

In order to perturb the classical action around classical trajectories, it is natural to perform the redefinition $X = B(X_{\text{cl}},Y)$, \i{i.e.}~$g=g_{\text{cl}}h$ in terms of $g=\exp(X)$ and $h=\exp(Y)$. Doing so in the action \eqref{euler_arnold_action} yields
\begin{equation}
\begin{split}
        S[g_{\text{cl}}h]= \int_0^T\dd t\Big(&
        \frac{1}{2}\langle I  \dot g_{\text{cl}}g_{\text{cl}} ^{-1},  \dot g_{\text{cl}}g_{\text{cl}} ^{-1}\rangle \\
        &+ 
        \langle I \dot g_{\text{cl}}g_{\text{cl}} ^{-1} , g_{\text{cl}}\dot h h ^{-1} g_{\text{cl}}^{-1}\rangle 
        +
        \frac{1}{2}\langle I g_{\text{cl}} \dot hh ^{-1} g_{\text{cl}}^{-1}, g_{\text{cl}} \dot h h ^{-1}g_{\text{cl}}^{-1}\rangle\Big).
\end{split}
        \label{330}
\end{equation}
There is a clear separation here between the first and second lines. The former is $S[g_{\text{cl}}]$, the action \eqref{euler_arnold_action} evaluated on the classical path, with a value \eqref{action is length square}. The latter consists of perturbations, labelled by $h(t)=\exp(Y(t))$, around the classical action.

An issue with the change of variables $g=g_{\text{cl}}h$ is that it amounts to a \i{left} multiplication (by $g_{\text{cl}}(t)$) at each time step, so it does not leave invariant the right path integral measure \eqref{pamelag}. This can be circumvented by trading left and right Haar measures through the modular function \eqref{mofo}. The path integral \eqref{euea} thus becomes
\begin{align}
        (g_f| e^{-i\hat H T}|e)
        = 
        \mathcal N\sum_{g_{\text{cl}}: e\overset{T}{\rightarrow} g_f}
        \left(\prod_{0<t<T} \frac{1}{\Delta(g_{\text{cl}}(t))}\right)
        \int_{Y(0)= 0}^{Y(T)=0}
        \mathcal{D}Y_{\text{R}}
        \;
        e^{iS[B(X_{\text{cl}},Y)]},
        \label{e59 ch'nord}
\end{align}
where the action functional $S[B(X_{\text{cl}},Y)]=S[g_{\text{cl}}h]$ is given by \eqref{330}. The sum runs over all classical paths $g_{\text{cl}}(t)$ that solve the Euler-Arnold equations of motion \eqref{velmom}--\eqref{euler_arnold_eom} and connect the identity $e$ to $g_f$ in time $T$.

Before expanding \eqref{e59 ch'nord} in powers of $Y$, let us rewrite the action \eqref{330} in a more convenient way. The left term in the bottom half of \eqref{330} can be rearranged by acting with the coadjoint action on the left side of the bracket, rather than the adjoint on the right. This produces the conserved charge \eqref{momentmap}. Similarly, in the last term of \eqref{330}, one can transfer the adjoint action to the left side of the bracket, leading to the definition of a `comoving inertia tensor'
\begin{align}
         I_{\text{cl}}(t) 
         :=
         \text{Ad}^*_{g_{\text{cl}}(t)^{-1}}\circ I\circ \text{Ad}_{g_{\text{cl}}(t)}
         \label{e60},
\end{align}
so that the action \eqref{330} can be recast as
\begin{align}
        S 
        =
        \frac{L[g_{\text{cl}}]^2}{2T} + \int_0^T \dd t\;\left(\langle Q, \dot hh ^{-1}\rangle + \frac{1}{2}\langle I_{\text{cl}}(t) \dot hh ^{-1}, \dot hh ^{-1}\rangle\right).
        \label{e61}
\end{align}
Note that, contrary to the factorized case of a free particle (see appendix \ref{app:U1}), the perturbation of the action depends on the classical path $g_{\text{cl}}(t)$ through the charge \eqref{momentmap} and the comoving inertia \eqref{e60}. Plugging \eqref{e61} in the propagator \eqref{e59 ch'nord}, one has
\begin{equation}
\begin{split}
        &(g_f| e^{-i\hat H T}|e)
        =\\
        &\mathcal{N}\!\!\!\!\sum_{g_{\text{cl}}:e\overset{T}{\rightarrow} g_f}\!\!\!e^{\frac{iL[g_{\text{cl}}]^2}{2T}} 
        \left(\prod_{0<t<T} \frac{1}{\Delta(g_{\text{cl}}(t))}\!\right)\!
        \int\limits_{Y(0)=0}^{Y(T)=0}\!\!\!\!\mathcal DY_{\text{R}}
        \exp\bigg[i\int_0^T\!\!\!\dd t \left(\langle Q, \dot hh ^{-1}\rangle + \frac{1}{2}\langle I_{\text{cl}}(t) \dot hh ^{-1}, \dot hh ^{-1}\rangle\right)\bigg].
        \end{split}
        \label{341}
\end{equation}
Our only remaining task is to evaluate perturbatively the path integrals in the summand of this expression.

\paragraph{Handling Haar measures.} The non-Euclidean measure on $G$ affects the evaluation of the path integral \eqref{341} in the semiclassical limit. Indeed, the saddle point approximation requires a flat measure, while our measure \eqref{pamelag} involves nontrivial Jacobian factors. The solution is to exponentiate the Jacobians, viewing them as an extra contribution to the action. We will do this using a Faddeev-Popov trick: since the Jacobians in \eqref{pamelag} are determinants of the right Maurer-Cartan form (see eq.~\eqref{e83} in appendix \ref{app:B}), one has
\begin{align}
    \int_{Y(0)=0}^{Y(T)=0}\mathcal{D}Y_{\text{R}}
    =
    \int_{Y(0)=0}^{Y(T)=0} \mathcal{D}Y\mathcal{D}\bar b\mathcal Db\; \exp\bigg[-\int_0^T \dd t \,\langle \bar b,\Omega^{\text{R}}_Y(b)\rangle\bigg],
    \label{e66}
\end{align}
where $\bar b$ and $b$ are Grassman-odd ghost fields, respectively valued in $\mathfrak{g}^*$ and $\mathfrak{g}$. The measure $\mathcal{D}Y$ is `flat', \i{i.e.}~it is given by eq.~\eqref{pamelag} \i{without} any Jacobian factor.

Upon plugging eq.~\eqref{e66} in the path integral \eqref{341}, the complete action, including ghosts, reads $W:=\int_0^T\dd t\,\mathcal W$ with a Lagrangian
\begin{equation}
        \mathcal W =
              \left\langle Q, \frac{e^{\text{ad}_{Y}}-1}{\text{ad}_{Y}}\dot Y \right\rangle
        +\frac{1}{2}
        \left\langle 
              I_{\text{cl}}(t)\frac{e^{\text{ad}_{Y}}-1}{\text{ad}_{Y}}\dot Y ,\frac{e^{\text{ad}_{Y}}-1}{\text{ad}_{Y}}\dot Y 
        \right\rangle
        +
        i \hbar \left\langle \bar b, \frac{e^{\text{ad}_{Y}}-1}{\text{ad}_{Y}}b\right\rangle,
        \label{e61bb}
\end{equation}
where we expressed the (right) Maurer-Cartan form in exponential coordinates (see eq.~\eqref{e71} in appendix \ref{app:B}). Note that we briefly reinstate $\hbar$, being understood that the integrand of the path integral is $e^{\frac{i}{\hbar}\int \mathcal W}$. The $\hbar$ factor in the last term of \eqref{e61bb} shows that measure terms are always quantum corrections to the classical action; indeed, they only start playing a role at two loops---see section \ref{ssecTwoLoop}.

\subsection{One-loop computation}
\label{ssec:oneloop}

Contrary to the case of a free particle on an Abelian group, the Lagrangian \eqref{e61bb} is by no means quadratic in $Y$. This is not an issue for the semiclassical regime, where one expands \eqref{e61bb} in powers of $Y$ anyway. The expansion can, in principle, be carried out to any order in $Y$. Here we focus on the first terms, which are quadratic in $Y$, corresponding to a one-loop calculation. The two-loop case is deferred to section \ref{ssecTwoLoop}.

\paragraph{Quadratic action.} The part of \eqref{e61bb} that is quadratic in $Y$ can be found thanks to the expansion of the Maurer-Cartan form near the identity (see eqs.~\eqref{e62}--\eqref{e63} in appendix \ref{app:B}). The ghosts do not contribute at that order---they can be integrated out---since the quadratic part of their action in \eqref{e66} is independent of $Y$. As a result, the quadratic part of the Lagrangian \eqref{e61bb}, without ghosts, reads
\begin{align}
        \mathcal W_0 
        =
        \frac{1}{2}\langle Q,[Y,\dot Y]\rangle + \frac{1}{2}\langle I_{\text{cl}}(t)\dot Y, \dot Y\rangle
        =
        \frac{1}{2}Q_{k}c^k{}_{ij}Y^i\dot Y^j
              +
        \frac{1}{2} I_{\text{cl}}(t)_{ij}\dot Y^i\dot Y^j,
        \label{e66bb}
\end{align}
where we also used the fact that \eqref{momentmap} is a conserved charge of the classical equations of motion. The corresponding quadratic action reads $W_0:=\int_0^T\dd t\,\mathcal W_0=\int_0^T \dd t \; Y^i\left(D_{ij}Y^j\right)$ in terms of the differential operator
\begin{align}
        D_{ij}(t) 
        :=
        Q_k c^k{}_{ij}\partial_t - \partial_t (I_{\text{cl}}(t)_{ij}\partial_t).
        \label{e339}
\end{align}
The one-loop approximation of each path integral in eq.~\eqref{341} is thus the (inverse of the square root of the) determinant of $D_{ij}$. Note that each such operator depends on the classical path $g_{\text{cl}}(t)$, both through the charge \eqref{momentmap} and through the comoving inertia \eqref{e60}.

\paragraph{Green's function.} Our strategy is to first compute the inverse of $D_{ij}$, then use it to deduce the sought-for determinant. A minor hassle is that both indices in \eqref{e339} are low, whereas only endomorphisms, strictly speaking, admit a determinant. We therefore raise one index in $D_{ij}$ by multiplication with the inverse of the comoving inertia \eqref{e60}, which produces the differential operator
\begin{align}
        (I_{\text{cl}}^{-1}D)^i{}_j 
        :=
        (I_{\text{cl}}(t)^{-1})^{il}D_{lj} 
        =
        Q_k c^k{}_{lj}(I_{\text{cl}}(t)^{-1})^{li}\partial_t - (I_{\text{cl}}(t)^{-1})^{ik}\dot I_{\text{cl}}(t)_{kj}\partial_t - \delta^i_j \partial_t^2.
        \label{e66b}
\end{align}
Note that, in contrast to \eqref{e339}, the second time derivative now appears without any time-dependent coefficient; this will be a key simplification. To streamline the notation, introduce the matrix
\begin{equation}
\label{s22}
A^i{}_j(t) 
=
-Q_k c^k{}_{lj}(I_{\text{cl}}(t)^{-1})^{li} +(I_{\text{cl}}(t)^{-1})^{ik}\dot I_{\text{cl}}(t)_{kj}
\end{equation}
so that eq.~\eqref{e66b} becomes
\begin{align}
    (I_{\text{cl}}^{-1}D)^i{}_j 
    =
    -A^i{}_j \partial_t -\delta^i_j \partial_t^2 
    =
    -\big[(\partial_t +A)\circ \partial_t\big]^i{}_j.
    \label{341b}
\end{align}
(We sometimes suppress Lie algebra indices $i,j$ for matrix-valued objects.) If $A=0$, the determinant of $I_{\text{cl}}^{-1}D$ reduces to the one relevant for a free particle. The matrix \eqref{s22} measures `how far' one is from the free, flat case.

The Green's function of $I_{\text{cl}} ^{-1} D$ is the distribution $G(t,s)$ that satisfies
\begin{align}
        -\partial_{t}^2 G(t,s)^i {}_{j} - A(t)^i {}_{k}\partial_t G^k {}_{j}(t,s) = \delta^i_j\delta(t-s),
        \label{e68b}
\end{align}
with $A$ defined in \eqref{s22} and Dirichlet-Dirichlet boundary conditions $G(0,s)=G(T,s)=0$. Following \cite{Gelfand:1959nq,Forman:1987gha}, we solve this by gluing together two linearly independent solutions of the differential equation $I_{\text{cl}} ^{-1} D\cdot f=0$, namely
\begin{align}
        M(t) := \int_0^t\dd s\,E(s)
        \quad\text{and}\quad
        N(t):= \int_t^T\dd s\,E(s),
        \quad\text{with}\quad
        E(t):=\mathcal{T}e^{-\int_0^t \dd\tau \; A(\tau)},
        \label{e72}
\end{align}
where $\mathcal{T}$ denotes time ordering, so that $\dot E + AE = 0$. These functions satisfy the boundary conditions $M(0)=0$ and $N(T)= 0$. The Green's function solving \eqref{e68b} is obtained by patching them together:
\begin{align}
G(t,s) = G_<(t,s) \Theta(s-t) + G_>(t,s)\Theta(t-s),
\label{s23}
\end{align}
where $\Theta(x):=\int_{-\infty}^x\delta(y)\dd y$ is the Heaviside theta function, and we let
\begin{equation}
G_<(t,s) := M(t) M(T)^{-1}  N(s) E(s)^{-1}
\quad\text{and}\quad
G_>(t,s) := N(t) M(T)^{-1} M(s) E(s)^{-1}.
\label{e519}
\end{equation}
The gluing \eqref{s23} makes the Green's function continuous, with a suitably discontinuous first derivative at $t=s$, ensuring that eq.~\eqref{e68b} holds along with Dirichlet-Dirichlet boundary conditions.

The Green's function \eqref{s23} yields correlation functions of the $Y$ variable defined above eq.~\eqref{330}, in the quadratic limit where the Lagrangian is \eqref{e66bb}. Indeed, note first that the Green's function is closely related tothe functional inverse of the differential operator \eqref{e66b}, since $D ^{-1}(t,s)^{ij} = G^{i}{}_{k}(t,s) (I_{\text{cl}}(s)^{-1})^{kj}$. Now define the free correlator
\begin{align}
         \langle X \rangle 
         :=
         \frac{\int_{Y(0)=0}^{Y(T)=0}\mathcal{D}Y\; X\, e^{i\int_0^T \frac{1}{2}\langle Y, D Y\rangle}}{\int_{Y(0)=0}^{Y(T)=0}\mathcal{D}Y\; e^{i\int_0^T \frac{1}{2}\langle Y, D Y\rangle}}
         \label{e360}
\end{align}
for any string $X=Y(t_1)\dots Y(t_K)$ of $Y$ variables, with $\mathcal{D}Y$ the same flat measure as on the right-hand side of \eqref{e66}. In particular, the free two-point correlator of $Y$ is
\begin{align}
         \langle Y^i(t) Y^j(s)\rangle
         =
         i D ^{-1}(t,s)^{ij} = i G^{i}{}_{k}(t,s) (I_{\text{cl}}(s)^{-1})^{kj}.
         \label{e361}
\end{align}
This is, as announced, the Green's function \eqref{s23} up to a factor $I_{\text{cl}}^{-1}$. Higher-point free correlators \eqref{e360} are given by products of such two-point correlators through Wick's theorem; this is essential for perturbation theory, and we return to it in section \ref{ssecTwoLoop}.

\paragraph{Functional determinant.} As usual with path integrals, only the ratio of a functional determinant with a suitable `reference' determinant can be evaluated. The natural reference, in the case at hand, is
\begin{equation}
\label{t23}
\operatorname{Det}(-\mathbb{I} \partial_t^2) 
=
\operatorname{Det}(-\partial_t^2)^n 
:=
(2\pi i T)^{n},
\end{equation}
where $\mathbb{I}$ is the $n\times n$ identity matrix and $-\partial_t^2$ acts on the space of functions of $t\in[0,T]$ that vanish at $t=0$ and $t=T$ (see appendix \ref{app:U1}). To compare the determinant of \eqref{341b} to \eqref{t23}, introduce the parametrized differential operator
\begin{align}
       I_{\text{cl}}^{-1} D_\lambda
       :=
       -\lambda A \partial_t - \partial_t^2
       \label{346}
\end{align}
that interpolates between $-\mathbb{I}\partial_t^2$ at $\lambda = 0$ and our operator \eqref{341b} at $\lambda = 1$. Also define the time-ordered exponential 
\begin{equation}
\label{ss23}
E_\lambda(t): = \mathcal T e^{-\lambda\int_0^t A(t)},
\end{equation}
the corresponding primitives $M_\lambda$ and $N_\lambda$ as in eqs.~\eqref{e72}, and the corresponding Green's function $G_\lambda(t,s)$ as in \eqref{s23}.

The determinant of \eqref{e339} can be extracted from its Green's function through a trace-log formula applied to the parametrized operator \eqref{346}. Explicitly, one has
\begin{align}
        \partial_\lambda \log(\operatorname{Det} I_{\text{cl}}^{-1}D_\lambda) 
        =
        -\Tr(G_\lambda A\partial_t) 
        =
        -\int_0^T\dd s \tr\Big(A(s) \frac{\partial}{\partial t}\Big|_{t=s}G_\lambda(t,s)\Big),
        \label{e74}
\end{align}
where $\Tr$ denotes the functional trace and $\tr$ is the matrix trace over Lie algebra indices. (Similarly, functional and matrix determinants will respectively be written $\text{Det}$ and $\det$.) Our goal is to write the right-hand side of \eqref{e74} as a total derivative in $\lambda$, then integrate from $\lambda=0$ to $\lambda =1$.

A subtlety is that the derivative of the Green's function \eqref{s23} is discontinuous at $t=s$, so a choice must be made to define what is meant by $\partial_t\big|_{t=s}G_\lambda(t,s)$ in \eqref{e74}. The answer stems from the discretization \eqref{e38}. Indeed, the Green's function \eqref{s23} is related to the two-point correlator \eqref{e361} of $Y$, which implies
\begin{equation}
\left\langle \tfrac{Y_{k+1}^i-Y_k^i}{\Delta T}Y_k^j\right\rangle
    =
    \langle \dot Y^i(t)Y^j(t)\rangle 
    =
    i\partial_t G^i{}_{k}(t,s)\big|_{t=s}I^{-1}_{\text{cl}}(t)^{kj},
    \label{tt21}
\end{equation}    
where we momentarily reinstate the definition of the discretized derivative given below \eqref{4.8}. The variable $Y_{k+1}^j$ appears strictly \i{after} $Y_k^j$, so only the positive branch $G_>(t,s)$ of eq.~\eqref{s23} must appear in the regularization of the equal-time derivative of the Green's function. In other words, the derivative of $G_{\lambda}$ in eq.~\eqref{e74} is to be understood as
$\partial_t\big|_{t=s}G_\lambda(t,s)=\partial_{t}\big|_{t=s} G_{>\lambda}(t,s) = E_\lambda(t)M_\lambda(T)^{-1}N_\lambda(t)E_\lambda(t)^{-1}-\mathbb I$, where we used \eqref{e519} and the property $M(t)+N(t)=M(T)$ that follows from \eqref{e72}. Plugging this back into \eqref{e74} and using the fact that $E_{\lambda}=-\dot N_{\lambda}=\dot M_{\lambda}$ is a parametrized time-ordered exponential \eqref{ss23} gives $\partial_\lambda \log(\operatorname{Det}I ^{-1} D_\lambda) = \partial_\lambda \log(\det M_\lambda(T)) + \int_0^T \dd t \tr A(t)$. This achieves our goal, stated below \eqref{e74}: its integral from $\lambda=0$ to 1 yields
\begin{align}
        \frac{\operatorname{Det}(I_{\text{cl}} ^{-1} D)}{\operatorname{Det}(-\mathbb{I}\partial_t^2)}
        =
        e^{\int_0^T\dd t \tr(A(t))}\frac{\det M(T)}{T^n},
        \label{e82}
\end{align}
where we used \eqref{t23} and $M_0(t) = t\,\mathbb{I}$ (set $A=0$ in \eqref{e72}).

The contribution of $A(t)$ in \eqref{e82} can be simplified further: the definition \eqref{s22} allows us to write $\int_0^T\dd t \tr(A(t)) 
         =
         \int_0^T \dd t \big({-}Q_k c^k{}_{ij}(I_{\text{cl}}(t)^{-1})^{ij} + \tr\big(I_{\text{cl}}(t)^{-1}\dot I_{\text{cl}}(t)\big)\big)$,
where the first term in the integrand vanishes because the symmetric tensor $I_{\text{cl}}(t)^{-1}$ is contracted with the antisymmetric structure constants of eq.~\eqref{struc}. As for the second term, it is the time derivative of $\tr(\log(I_{\text{cl}})) = \log(\det I_{\text{cl}})$, so
\begin{align}
    e^{\int_0^T\dd t \tr(A(t))}
    = 
    \frac{\det I_{\text{cl}}(T)}{\det I_{\text{cl}}(0)}
    =
    \frac{1}{\Delta(g_f)^2}.
    \label{eab}
\end{align}
Here we used $g(t=0)=g_i=e$, $g(t=T)=g_f$ and the definition \eqref{e60} of $I_{\text{cl}}(t)$ to identify the modular function $\Delta(g)=1/\det(\text{Ad}_g)$ (see appendix \ref{app:B}) through
\begin{align}
    \det(I_{\text{cl}}(t)) = \frac{\det(I)}{\Delta(g_{\text{cl}}(t))^2}.
    \label{edetIcl}
\end{align}
The final result for the determinant \eqref{e82} is thus
\begin{align}
\frac{\operatorname{Det}(I_{\text{cl}} ^{-1} D)}{\operatorname{Det}(-\mathbb{I}\partial_t^2)}
=
\frac{\det(M(T))}{T^n\Delta(g_f)^2},
\qquad\text{hence}\qquad
\operatorname{Det}(I_{\text{cl}} ^{-1} D)
=
(2\pi i)^n\frac{\det(M(T))}{\Delta(g_f)^2},
\label{e85}
\end{align}
where we recall that $M(t)$ was defined in eq.~\eqref{e72} and we used \eqref{t23} for the reference determinant.

\paragraph{One-loop propagator.} What is actually needed in the path integral \eqref{341} is the functional determinant of $D$ itself, $\operatorname{Det}(D) = \operatorname{Det}(I_{\text{cl}})\operatorname{Det}(I_{\text{cl}} ^{-1} D )$. Using \eqref{edetIcl}, one formally has
\begin{equation}
\operatorname{Det}(I_{\text{cl}}) = \prod_{0<t<T} \det(I_{\text{cl}}(t)) = \prod_{0<t<T}\det(I)\prod_{0<t<T}\frac{1}{\Delta(g_{\text{cl}}(t))^2}.
\label{e530}
\end{equation}
Gaussian integration over $Y$ in \eqref{341}, with the quadratic Lagrangian \eqref{e66bb}, leads to a $1/\sqrt{\operatorname{Det}(I_{\text{cl}})}$ contribution, which cancels the product of modular functions in \eqref{341}. Moreover, the product of determinants of $I$ in \eqref{e530} cancels almost entirely the product \eqref{s16} that stems from the integral over momenta. The only factor that remains is the $\sqrt{\det I}$ that stems from integration over the momentum at time $t=0$, since the discretized path integral \eqref{e44} contains one more identity in momentum space than in position space. At the end of the day, the one-loop approximation of the propagator \eqref{341} reads
\begin{align}
         (g_f| e^{-i\hat H T}|e)
         \sim
         \Delta(g_f)\sqrt{\frac{\det I}{(2\pi i)^{n}}}\sum_{g_{\text{cl}}:e\overset{T}{\rightarrow} g_f} \frac{e^{\frac{iL[g_{\text{cl}}]^2}{2T}}}{\sqrt{\det M[g_{\text{cl}}](T)}},
         \label{e89}
\end{align}
where we further stress the dependence of $M(T)$, defined in \eqref{e72}, on the classical path $g_{\text{cl}}(t)$. The latter solves the Euler-Arnold equations of motion \eqref{velmom}--\eqref{euler_arnold_eom} in such a way that $g_{\text{cl}}(0)=e$ and $g_{\text{cl}}(T)=g_f$.

The time dependence of the propagator \eqref{e89} can actually be made explicit, thanks to the scaling of solutions of the Euler-Arnold equations of motion discussed below \eqref{action is length square}. Indeed, let $g_{\text{ref}}$ be the geodesic path of length $L$ that connects $g_i$ to $g_f$ in time $1$, so that $g_{\text{cl}}(t) = g_{\text{ref}}(t/T)$. This leads to a definite scaling of the matrix \eqref{s22}, namely $A(t) = \frac{1}{T} A_{\text{ref}}(t/T)$, hence $\det M(T) = T^n \det M_{\text{ref}}(1)$ for the matrix defined in \eqref{e72}. The propagator \eqref{e89} can finally be recast as
\begin{align}
         (g_f| e^{-i\hat H T}|e)
         =
         \Delta(g_f)\sqrt{\frac{\det I}{(2\pi i T)^n}}\sum_{g_{\text{ref}}:e\overset{1}{\rightarrow} g_f} \frac{e^{\frac{iL[g_{\text{ref}}]^2}{2T}}}{\sqrt{\det M_{\text{ref}}[g_{\text{ref}}](1)}} ,
         \label{final 1 loop propagator}
\end{align}
where the paths appearing in the sum are only those classical solutions that connect the identity to $g_f$ in unit time.

\subsection{Two-loop computation}
\label{ssecTwoLoop}

Perturbative computations of path integrals are prone to divergences due to the functional nature of the integrand. A key sanity check for eq.~\eqref{341} is that such divergences cancel out, eventually giving a finite result. Here, we prove this cancellation at two-loop order, for any Euler-Arnold system. The expression of the resulting two-loop propagator is rather cumbersome, so its details are provided in appendix \ref{app:2loop}.

\paragraph{Short-time expansion.} The propagator \eqref{341} is a heat kernel on $G$ for any Euler-Arnold Hamiltonian \eqref{euler-arnold}, so its semiclassical (loop) expansion is actually a short-time expansion given by the De Witt ansatz (see \i{e.g.}~\cite[sec.~4.3]{Vassilevich:2003xt}). In that context, the one-loop formula \eqref{final 1 loop propagator} is but the leading term of an asymptotic series
\begin{align}
    (g_f|e^{-i\hat H T}|e) 
    =
    \Delta(g_f)\sqrt{\frac{\det I}{2\pi i T^n}}\!\!\!\sum_{g_{\text{ref}}:e\overset{1}{\rightarrow} g_f}\!\!\!
    \frac{e^{\frac{iL[g_{\text{ref}}]^2}{2T}}}{\sqrt{\det M_{\text{ref}}[g_{\text{ref}}](1)}}
    \big(1 + iT b_2[g_{\text{ref}}] + (iT)^2 b_3[g_{\text{ref}}] +\dots\big),
    \label{dewitt}
\end{align}
where each $b_m$ results from an $m$-loop calculation. The one-loop determinant $\frac{\Delta(g_f)^2\det(I)}{\det(M_{\text{ref}}[g_{\text{ref}}](1))}$ is then known as the Van Vleck-Morette determinant \cite{DeWitt-Morette}.

Our focus here is on the two-loop term $b_2$ in the expansion \eqref{dewitt}. Its expression is found with standard perturbation theory: write the Lagrangian \eqref{e61bb} as a power series
\begin{align}
    \mathcal W = \mathcal W_0 + \mathcal W_3 + \mathcal W_4 + O(Y^5),
    \label{e535}
\end{align}
and expand the path integral in \eqref{341} as
\begin{align}
    \frac{\int_{Y(0)=0}^{Y(T)=0}\mathcal DY\mathcal D\bar b\mathcal Db \exp(iW)}{\int_{Y(0)=0}^{Y(T)=0}\mathcal DY\mathcal D\bar b\mathcal Db \exp(iW_0)} ]
    \sim
     1+ i\int_0^T\dd t \langle \mathcal W_4(t)\rangle - \frac{1}{2}\int_0^T\dd t \int_0^T\dd s \langle \mathcal W_3(t)\mathcal W_3(s)\rangle.
    \label{5.39TheHourOfWritingStuff}
\end{align}
Here we respectively denote the action and its quadratic approximation by $\mathcal W$ and $\mathcal W_0$, as defined above eqs.~\eqref{e61bb} and \eqref{e339}, and expectation values are defined as in \eqref{e360}.

The right-hand side of \eqref{5.39TheHourOfWritingStuff} can be written as $1+iTb_2$ in terms of the De Witt ansatz \eqref{dewitt}. It manifestly involves correlators of the variables $Y$ and the ghosts $b,\bar b$ introduced in \eqref{e66}. The corresponding two-point functions are \eqref{e361} and
\begin{align}
    \langle b^\alpha(t) \bar b_\beta(s)\rangle = \delta^\alpha_\beta\delta(t-s),
    \label{ghost2pt}
\end{align}
while the two-point functions $\langle Yb\rangle=\langle Y\bar b\rangle=0$ vanish. These two-point functions determine all correlators needed for perturbation theory, by Wick's theorem.

A remark on notation: to simplify the diagrammatics until the end of this section, Latin and Greek letters both denote Lie algebra indices, but we devote the beginning of the Latin alphabet to contractions with $Y$, and Greek indices for contractions with ghosts. Indices $i,j,k$ from the middle of the Latin alphabet are devoted to internal contractions of Lie algebra indices. 

\paragraph{Vertices and divergences.} For the Euler-Arnold Lagrangian \eqref{e61bb}, both $\mathcal W_3$ and $\mathcal W_4$ (and any $\mathcal W_n$ in the expansion \eqref{e535}) contain a term with one $\dot Y$, a term with two $\dot Y$s and a term with two ghosts. The expansion \eqref{e63} of the right Maurer-Cartan form explicitly yields
\begin{align}
    \mathcal W_3 = V^{3Y1D}_{a(bc)}\dot Y^aY^bY^c + V^{3Y2D}_{(ab)c}\dot Y^a\dot Y^b Y^c + V^{1Y2G}{}^\alpha{}_{\beta a} \bar b_\alpha b^\beta Y^a
    \label{W3}
\end{align}
for the cubic part of \eqref{e535}, with the vertices
\begin{align}
    &V^{3Y1D}_{a(bc)} = -\frac{1}{6}c^j{}_{a(b}c^i{}_{c)j}Q_i, &\text{denoted} & &
\begin{tikzpicture}
    \coordinate (V) at (0,0);
    \draw[matter] (V) -- (0,0.7);
    \draw[matter] (V) -- (210:0.7);
    \draw[matter] (V) -- (330:0.7);
    \node[dot] at (0,0.25) {};
\end{tikzpicture},
\label{3Y1D}
    \\
    &V^{3Y2D}_{(ab)c} = -\frac{1}{2}c_{(ab)c}, &\text{denoted} & &
    \begin{tikzpicture}
    \coordinate (V) at (0,0);
    \draw[matter] (V) -- (0,0.7);
    \draw[matter] (V) -- (210:0.7);
    \draw[matter] (V) -- (330:0.7);
    \node[dot] at (210:0.25) {};
    \node[dot] at (330:0.25) {};
\end{tikzpicture},
\label{3Y2D}
    \\
    &V^{1Y2G}{}^\alpha{}_{\beta a} = -\frac{i}{2}c^{\alpha}{}_{\beta a}, &\text{denoted} & &
    \begin{tikzpicture}
    \coordinate (V) at (0,0);
    \draw[matter] (V) -- (0,0.7);
    \draw[ghost] (V) -- (210:0.7);
    \draw[ghost] (V) -- (330:0.7);
\end{tikzpicture}.
\label{1Y2G}
\end{align}
Parentheses enclosing indices denote symmetrization as in $T_{(ab)}=\frac{1}{2}(T_{ab}+T_{ba})$; indices are raised and lowered thanks to the comoving inertia \eqref{e60} and its inverse. One must therefore keep in mind that $c_{ijk}(t):= I_{\text{cl}}(t)_{il}c^l{}_{jk}$ depends on time even though the structure constants in \eqref{struc} are time-independent. In the diagrammatic representation \eqref{3Y1D}--\eqref{1Y2G}, solid lines emanating from a vertex represent a contraction with a $Y$, while solid lines with a black dot denote a contraction with a $\dot Y$. Dotted lines denote contractions with a ghost. 

Similarly, pushing the expansion \eqref{e63} of Maurer-Cartan forms to the next order yields the quartic term in \eqref{e535},
\begin{align}
    \mathcal W_4 = V^{4Y1D}_{a(bcd)}\dot Y^a Y^b Y^c Y^d + V^{4Y2D}_{(ab)(cd)}\dot Y^a \dot Y^b Y^cY^d + V^{2Y2G}{}^\alpha{}_{\beta(ab)} \bar b_\alpha b^\beta Y^a Y^b,
    \label{W4}
\end{align}
with vertices now given by
\begin{align}
    &V^{4Y1D}_{a(bcd)} = -\frac{1}{24}c^k{}_{a(b}c^j{}_{c|k|}c^i{}_{d)j}Q_i,
    &\text{denoted} & &
    \begin{tikzpicture}
    \coordinate (V) at (0,0);
    \draw[matter] (V) -- (45:0.7);
    \draw[matter] (V) -- (135:0.7);
    \draw[matter] (V) -- (225:0.7);
    \draw[matter] (V) -- (315:0.7);
    \node[dot] at (45:0.25) {};
\end{tikzpicture},
\label{4Y1D}
    \\
    &V^{4Y2D}_{(ab)(cd)}=-\frac{1}{8}c^i{}_{c)(a}c_{|i|b)(d}+\frac{1}{6}c_{a)(c|j|}c^{j}{}_{d)(b}, & \text{denoted} & &
    \begin{tikzpicture}
    \coordinate (V) at (0,0);
    \draw[matter] (V) -- (45:0.7);
    \draw[matter] (V) -- (135:0.7);
    \draw[matter] (V) -- (225:0.7);
    \draw[matter] (V) -- (315:0.7);
    \node[dot] at (45:0.25) {};
    \node[dot] at (135:0.25) {};
\end{tikzpicture},
\label{4Y2D}
    \\
    &V^{2Y2G}{}^{\alpha}{}_{\beta(ab)} = -\frac{i}{6}c^{\alpha}{}_{i(a}c^i{}_{b)\beta}, &\text{denoted} & &
    \begin{tikzpicture}
    \coordinate (V) at (0,0);
    \draw[matter] (V) -- (45:0.7);
    \draw[matter] (V) -- (135:0.7);
    \draw[ghost] (V) -- (225:0.7);
    \draw[ghost] (V) -- (315:0.7);
\end{tikzpicture}.
\label{2Y2G}
\end{align}
These vertices form the starting point of path-integral perturbation theory.

At this point, some regulating prescriptions remain to be clarified. Focus for definiteness on the vertex \eqref{4Y2D} appearing in the quartic term of \eqref{5.39TheHourOfWritingStuff}. The resulting average is
\begin{align}
    i\int_0^T \dd t \;V^{4Y2D}_{abcd} \langle \dot Y^a\dot Y^bY^cY^d\rangle &= i\int_0^T \dd t \;V^{4Y2D}_{abcd} \left(\langle \dot Y^a\dot Y^b\rangle \langle Y^cY^d\rangle+2\langle\dot Y^aY^c\rangle\langle\dot Y^bY^d\rangle\right),
    \label{547}
\end{align}
where we used Wick's theorem on the right-hand side. The two contractions on the right-hand side are graphically represented as
\begin{align}
    i\int_0^T \dd t \;V^{4Y2D}_{abcd} \langle \dot Y^a\dot Y^bY^cY^d\rangle = \begin{tikzpicture}
    \draw[matter] (0,0.55) circle (0.55);
    \draw[matter] (0,-0.55) circle (0.55);
    \node[dot] at ({-0.55*cos(50)}, {0.55 - 0.55*sin(50)}) {};
    \node[dot] at ({0.55*cos(50)}, {0.55 - 0.55*sin(50)}) {};
\end{tikzpicture} + 2 \;\begin{tikzpicture}
    \draw[matter] (0,0.55) circle (0.55);
    \draw[matter] (0,-0.55) circle (0.55);
    \node[dot] at ({-0.55*cos(50)}, {0.55 - 0.55*sin(50)}) {};
    \node[dot] at ({-0.55*cos(50)}, {-0.55 + 0.55*sin(50)}) {};
\end{tikzpicture}.
\label{549}
\end{align}
Given the definition \eqref{e361} of the two-point function, the second term on the right hand side involves derivatives of the Green function at equal times: $\langle \dot Y^a(t)Y^c(t)\rangle = i\partial_1D^{-1}(t,t)^{ac}$, where $\partial_1$ denotes the derivative with respect to the first argument. This was addressed around eq.~\eqref{tt21}, so the correct prescription is
\begin{align}
    \langle \dot Y^a(t)Y^c(t)\rangle= i\partial_1D_>^{-1}(t,t)^{ac}.
    \label{oneder}
\end{align}
However, the first term on the right-hand side of \eqref{549} contains bigger issues in $\langle \dot Y^a(t)\dot Y^b(t)\rangle = \partial_1\partial_2D^{-1}(t,t)^{ab}$, since second derivatives of the Green function contain delta distributions as in \eqref{e68b}. Taking two derivatives of \eqref{e361} leads to 
\begin{align}
    \partial_1\partial_2D^{-1}(t,s) = \delta(t-s) I_{\text{cl}}(s)^{-1} + \partial_1\partial_2D^{-1}_{\text{reg}}(t,s),
    \label{doubleder}
\end{align}
with 
\begin{align}
    \partial_1\partial_2D^{-1}_{\text{reg}}(t,s) := {-E(t)M(T)^{-1} I_{\text{cl}}(s)^{-1} + \partial_1D^{-1}(t,s) \text{ad}^*_{I_{\text{cl}}(s)^{-1}\cdot}Q},
    \label{doublederreg}
\end{align}
where we used matrix notations. The matrix $\text{ad}^*_{I_{\text{cl}}(s)^{-1}\cdot} Q$ is a tensor with one upper and one lower index that first turns an element of $\mathfrak{g}^*$ into an element of $\mathfrak{g}$ through contraction with $I_{\text{cl}}(s)^{-1}$, and then returns the coadjoint action of the resulting Lie algebra element on the conserved charge \eqref{momentmap}. In index notation, one has $(\text{ad}^*_{I_{\text{cl}}(s)^{-1}\cdot} Q)_i{}^j=-Q_lc^{l}{}_{ik}(I_{\text{cl}}(s)^{-1})^{kj}$.\footnote{Using the self-adjointness of the operator \eqref{e339}, the matrix $\text{ad}^*_{I_{\text{cl}}(s)^{-1}\cdot} Q$ can be expressed in terms of $E$ as $\text{ad}^*_{I_{\text{cl}}(s)^{-1}\cdot} Q = E(s)M^{-1}(s)(I^{-1}E(s)^TI_{\text{cl}}(s)-1)$, where $T$ denotes matrix transposition.}

 Trying to evaluate \eqref{doubleder} at equal times generates two issues. The first is a $\delta(0)$ divergence. As we prove in the remainder of this section, all such $\delta(0)$ divergences cancel each other out. The second is yet another equal time evaluation of the derivative of the propagator in the regular part \eqref{doublederreg}. This time, the prescription \eqref{oneder} does not apply. Indeed, going back to the discretized setup, one has
\begin{align}
    \langle \dot Y^a(t)\dot Y^b(t)\rangle = \left\langle \frac{Y_{k+1}^a-Y_k^a}{\Delta T} \frac{Y_{k+1}^b-Y_k^b}{\Delta T}\right\rangle.
    \label{discr}
\end{align}
Here the time steps $k+1$ and $k$ are on a symmetric footing; in fact, carefully evaluating the expectation value on the right-hand side of \eqref{discr} and taking the time step to zero leads to $
    \langle \dot Y^a(t)\dot Y^b(t)\rangle = I_{\text{cl}}^{-1}\delta(0) + \frac{1}{2}\left(\partial_1\partial_2D^{-1}_>(t,t)^{ab} + \partial_1\partial_2D^{-1}_<(t,t)^{ab}\right)
$. The prescription for \eqref{doubleder} thus becomes
\begin{align}
    \partial_1\partial_2D^{-1}(t,t) 
    &=
    \delta(0) I_{\text{cl}}(t)^{-1}-E(t)M(T)^{-1} I_{\text{cl}}(t)^{-1} + \Big(\partial_1D_>^{-1}(t,t)+\frac{1}{2}I_{\text{cl}}(t)^{-1}\Big) \text{ad}^*_{I_{\text{cl}}(t)^{-1}\cdot}Q,
    \label{554}
\end{align}
where we chose to express the regulation solely in terms of $\partial_1D_>^{-1}$ to make contact with the first derivative prescription \eqref{oneder}, with
\begin{align}
    \partial_1D^{-1}(t,s) = E(t)M(T)^{-1}(N(s)\Theta(s-t)-M(s)\Theta(t-s))E(s)^{-1}I_{\text{cl}}(s)^{-1}.
    \label{555}
\end{align}
We now know how to evaluate each diagram coming in the loop expansion.

\paragraph{Cancellation of divergences.} Divergences involving factors $\delta(0)$ arise from three distinct sources. As in the example \eqref{547}, they occur when contracting two one-derivative legs emanating from the same vertex together. Similarly, as deduced from the ghost two-point functions \eqref{ghost2pt}, they arise when contracting two ghost lines emanating from the same vertex, as in
\begin{align}
    \begin{tikzpicture}
    \draw[matter] (0,0.55) circle (0.55);
    \draw[ghost] (0,-0.55) circle (0.55);
\end{tikzpicture}\,.
\label{2loopV2Y2G}
\end{align}
Finally, they appear upon contracting two pairs of one-derivative legs or ghost legs emanating from two different vertices. Examples of such cases are
\begin{align}
\begin{tikzpicture}
    \draw[matter] (0,0) circle (0.75);
    \draw[matter] (0,0.75) -- (0,-0.75);
    \node[dot] at (70:0.75) {};
    \node[dot] at (110:0.75) {};
    \node[dot] at (-70:0.75) {};
    \node[dot] at (-110:0.75) {};
\end{tikzpicture}
\quad \text{and}\quad
\begin{tikzpicture}
    \draw[ghost] (0,0) circle (0.75);
    \draw[matter] (0,0.75) -- (0,-0.75);
\end{tikzpicture}\,.
\label{557}
\end{align}
These rules imply that all diagrams with ghosts contain divergences. When checking for the cancellation of divergences, a rule of thumb is that divergences coming from contractions of two one-derivative lines are usually cancelled by divergences coming from the same diagram with two one-derivative lines replaced by ghost lines. Consequently, one can show that the divergences coming from
\begin{align*}
    \begin{tikzpicture}
    \draw[matter] (0,0.75) circle (0.4);
    \draw[ghost] (0,-0.75) circle (0.4);
    \draw[matter] (0,0.35) -- (0,-0.35);
    \node[dot] at ({-0.4*cos(40)}, {0.75 - 0.4*sin(40)}) {};
\end{tikzpicture}
\quad \text{and} \quad
\begin{tikzpicture}
    \draw[matter] (0,0.75) circle (0.4);
    \draw[matter] (0,-0.75) circle (0.4);
    \draw[matter] (0,0.35) -- (0,-0.35);
    \node[dot] at ({-0.4*cos(40)}, {0.75 - 0.4*sin(40)}) {};
    \node[dot] at ({-0.4*cos(50)}, {-0.75 + 0.4*sin(50)}) {};
    \node[dot] at ({0.4*cos(50)}, {-0.75 + 0.4*sin(50)}) {};
\end{tikzpicture},
\qquad \qquad \qquad 
\begin{tikzpicture}
    \draw[matter] (0,0.75) circle (0.4);
    \draw[ghost] (0,-0.75) circle (0.4);
    \draw[matter] (0,0.35) -- (0,-0.35);
    \node[dot] at (0, 0.18) {};
\end{tikzpicture}
\quad \text{and} \quad
\begin{tikzpicture}
    \draw[matter] (0,0.75) circle (0.4);
    \draw[matter] (0,-0.75) circle (0.4);
    \draw[matter] (0,0.35) -- (0,-0.35);
    \node[dot] at (0, 0.18) {};
    \node[dot] at ({-0.4*cos(50)}, {-0.75 + 0.4*sin(50)}) {};
    \node[dot] at ({0.4*cos(50)}, {-0.75 + 0.4*sin(50)}) {};
\end{tikzpicture}
,\qquad \qquad \qquad
\begin{tikzpicture}
    \draw[matter] (0,0.75) circle (0.4);
    \draw[matter] (0,-0.75) circle (0.4);
    \draw[matter] (0,0.35) -- (0,-0.35);
    \node[dot] at ({-0.4*cos(50)}, {0.75 - 0.4*sin(50)}) {};
    \node[dot] at ({0.4*cos(50)}, {0.75 - 0.4*sin(50)}) {};
    \node[dot] at (0, -0.22) {};
    \node[dot] at ({0.4*cos(50)}, {-0.75 + 0.4*sin(50)}) {};
\end{tikzpicture}
\quad \text{and} \quad
\begin{tikzpicture}
    \draw[ghost] (0,0.75) circle (0.4);
    \draw[matter] (0,-0.75) circle (0.4);
    \draw[matter] (0,0.35) -- (0,-0.35);
    \node[dot] at (0, -0.22) {};
    \node[dot] at ({0.4*cos(50)}, {-0.75 + 0.4*sin(50)}) {};
\end{tikzpicture},
\end{align*}
as well as the divergences coming from the three diagrams
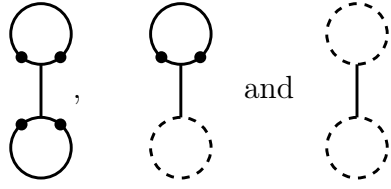
\begin{align*}
\begin{tikzpicture}
    \draw[matter] (0,0.75) circle (0.4);
    \draw[matter] (0,-0.75) circle (0.4);
    \draw[matter] (0,0.35) -- (0,-0.35);
    \node[dot] at ({-0.4*cos(50)}, {0.75 - 0.4*sin(50)}) {};
    \node[dot] at ({0.4*cos(50)}, {0.75 - 0.4*sin(50)}) {};
    \node[dot] at ({-0.4*cos(50)}, {-0.75 + 0.4*sin(50)}) {};
    \node[dot] at ({0.4*cos(50)}, {-0.75 + 0.4*sin(50)}) {};
\end{tikzpicture}, \quad \quad 
\begin{tikzpicture}
    \draw[matter] (0,0.75) circle (0.4);
    \draw[ghost] (0,-0.75) circle (0.4);
    \draw[matter] (0,0.35) -- (0,-0.35);
    \node[dot] at ({-0.4*cos(50)}, {0.75 - 0.4*sin(50)}) {};
    \node[dot] at ({0.4*cos(50)}, {0.75 - 0.4*sin(50)}) {};
\end{tikzpicture}
\quad \text{and} \quad 
\begin{tikzpicture}
    \draw[ghost] (0,0.75) circle (0.4);
    \draw[ghost] (0,-0.75) circle (0.4);
    \draw[matter] (0,0.35) -- (0,-0.35);
\end{tikzpicture}
\end{align*}
cancel each other out. Finally, in the divergence of the left diagram in \eqref{549}, only the $1/6$ part of the vertex \eqref{4Y2D} is cancelled by \eqref{2loopV2Y2G}, while the $1/8$ part of the vertex \eqref{4Y2D} cancels the divergences coming from the two diagrams in \eqref{557}.

Ultimately, each and every $\delta(0)$ is cancelled at two loops. This occurs thanks to the ghosts in \eqref{e66}, which account for the non-flatness of the measure. It seems reasonable to expect similar cancellations to occur at higher loop orders: this amounts to discarding all diagrams that contain ghosts, and all $\delta(0)$ terms that would arise from second derivatives \eqref{doubleder}. Mind that it does not mean that second derivatives \eqref{doubleder} can be entirely replaced by their regular counterpart \eqref{doublederreg}, since contractions between two different vertices can yield finite contributions from the non regular part of \eqref{doubleder}.


\section{Partition functions}
\label{sec:partition}

The partition function of a quantum system contains information on its equilibrium properties. Here, we study the partition functions of Euler-Arnold systems. We also consider characters of the left and right regular representations, which may be seen as partition functions in $L^2(G)$ with a linear (as opposed to quadratic) Lie-Poisson Hamiltonian.

\subsection{Partition functions of Euler-Arnold systems}
\label{ssec:partition Euler Arnold}

Let $\hat H$ be a Hamiltonian operator acting on $L^2(G)$. The corresponding canonical partition function at temperature $1/\beta$ is
\begin{align}
         Z(\beta)
         =
         \Tr \big(e^{-\beta \hat H}\big)
         =
         \int \dd g_{\text{L}} \left( g \right|e^{-\beta \hat H} \left| g \right).
         \label{e41}
\end{align}
If the Hamiltonian is invariant under right group translations, as is the case for Lie-Poisson systems, the integrand can be simplified thanks to an analogue of eq.~\eqref{e right invariance}. The partition function \eqref{e41} thus becomes
\begin{align}
          Z(\beta)
         =
         \int \dd g_{\text{L}}\; \Delta(g^{-1}) \left(e\right|e^{-i\hat H T}\left|e\right) 
          =
          \int \dd g_{\text{R}} \left(e\right|e^{-i\hat H T}\left|e\right) 
          =
          \operatorname{Vol}_{\text{R}}(G) \left(e\right|e^{-i\hat H T}\left|e\right)
         \label{e42},
\end{align}
where $\operatorname{Vol}_{\text{R}}(G):= \int \dd g_{\text{R}}$ is the volume of the group with respect to the right Haar measure. In case the volume diverges, one can still define the partition function \i{density} as the local average $z(\beta) := (e|e^{-\beta\hat H}|e)$.

Upon identifying $\beta = -iT$, this average becomes a Wick-rotated version of the propagator \eqref{e55}, so the tools of section \ref{sec:G} apply. There is just one caveat: the point $g_f$ in the Lorentzian path integral \eqref{e55} is assumed to have only discretely many logarithms \eqref{genlogs}, which is untrue for the identity element in \eqref{e42} (recall footnote \ref{foo}). To account for that, one should pick a fixed Cartan subalgebra $\mathfrak h$, and only sum over logarithms of the identity in this Cartan subalgebra. Then the partition function (density) reads
\begin{align}
\label{e63bb}
        z(\beta)
        =
        \mathcal N \sum_{Y\in \operatorname{Logs}(e)\cap \mathfrak{h}} \int_{X(0)= 0}^{X(\beta) = Y} \mathcal{D}X_{\text{R}} \,e^{-\int_0^\beta \dd \tau \frac{1}{2}\langle I \,\Omega^{\text{R}}_X(\dot X), \Omega^{\text{R}}_X(\dot X)\rangle},
\end{align}
where the dot now means a derivative with respect to imaginary time $\tau = it$.

We stress that the restriction to logarithms that are contained in a Cartan subalgebra is not a matter of mere convenience. It originates, instead, from the full group integral \eqref{e41}, \i{before} using right-invariance for the simplification \eqref{e42}. Indeed, almost all group elements in the integral \eqref{e41} are `generic', with logarithms given by \eqref{genlogs}, so the presence of an exceptional number of logarithms at the identity occurs on a set of measure zero. This leaves the full partition function \eqref{e41} unaffected, so its density \eqref{e63bb} must be evaluated as if the identity were a generic point in $G$, with countably many logarithms \eqref{genlogs}.

\paragraph{One-loop partition function.} Let us now focus on Euler-Arnold systems. Since there is no potential term in the Hamiltonian \eqref{euler-arnold}, the Euclidean action in the exponent of \eqref{e63bb} equals the real-time Lagrangian action \eqref{euler_arnold_action}. It follows that all computations regarding functional determinants and propagators still hold, almost unchanged. A minor exception is the reference determinant \eqref{t23}, whose Wick-rotated version is $\operatorname{Det}(D_0) = (2\pi \beta)^n$. Eq.~\eqref{final 1 loop propagator} then yields the following one-loop formula for the partition function density:
\begin{align}
         z(\beta)
         \sim
         \sqrt{\frac{\det I}{(2\pi\beta)^n}}\sum_{g_{\text{cl}}:g\overset{1}{\rightarrow} g} \frac{e^{-\frac{L[g_{\text{cl}}]^2}{2\beta}}}{\sqrt{\det M[g_{\text{cl}}](1)}}.
         \label{e44bb}
\end{align}
Note that there is no contribution of the modular function here, since $\Delta(e)=1$. In the high-temperature limit $\beta\rightarrow 0$, only the shortest path contributes in the sum over geodesic loops; it is the constant path $g_{\text{cl}}(\tau) = e$, of length $L[g_{\text{cl}}]=0$. In that case, everything trivializes: the comoving inertia \eqref{e60} reduces to $I_{\text{cl}}(\tau) = I$, the charge \eqref{momentmap} vanishes, as does the matrix \eqref{s22}. The end result is the familiar partition function density
\begin{align}
         z(\beta)
         \overset{\beta \rightarrow 0}{\sim}
         \sqrt{\frac{\det I}{(2\pi\beta)^n}}
         \label{e65bb}
\end{align}
of a free particle in dimension $n$, with mass tensor $I$, consistent with the equipartition theorem of classical mechanics. Note that, at high temperature, the contributions from paths of greater length are exponentially suppressed, so the only perturbative corrections to \eqref{e65bb} are higher-loop effects around the constant path. These can be computed perturbatively by applying the same method as in section \ref{ssecTwoLoop} around the trivial saddle point.

\paragraph{Two-loop partition function at high temperature.} Eq.~\eqref{e65bb} may be seen as the leading term of the high-temperature expansion of the partition function \eqref{e63bb}. Similarly to one-loop order, the identification $iT = \beta$ implies that the two-loop results of appendix \ref{app:2loop} apply to the partition function.

A tremendous simplification stems from the fact that the only saddle that contributes at high temperature is the constant path, $g(t)=e$ for all $t$. This makes it possible to compute the two-loop partition function in closed form. Indeed, following the simplifications detailed above eq.~\eqref{e65bb}, the one-derivative vertices \eqref{3Y1D} and \eqref{4Y1D} are proportional to $Q=0$, so they can be discarded and the only nonzero contributions to the two-loop partition function are \eqref{C3} and \eqref{C5}. Moreover, since the matrix \eqref{s22} vanishes for constant paths, the Green's function \eqref{e361} can be computed explicitly as
\begin{align}
    D^{-1}(t,s)^{ab} = (I^{-1})^{ab}\Big(\min(t,s) - \frac{ts}{T}\Big).
    \label{greenpartition}
\end{align}
Differentiating this yields all the tensors needed to evaluate \eqref{C3} and \eqref{C5}, namely: $\partial_1D^{-1}(t,s) = I^{-1}\left((1-\frac{s}{T})\Theta(s-t) - \frac{s}{T}\Theta(t-s)\right)$ and in particular $\partial_1D^{-1}_>(t,t) = -I^{-1}\frac{t}{T}$, as well as $\partial_1\partial_2D^{-1}(t,s) = I^{-1}\left(\delta(t-s)-\frac{1}{T}\right)$ and in particular $\partial_1\partial_2D^{-1}_{\text{reg}}=-\frac{I^{-1}}{T}$. Plugging that into \eqref{C3} and \eqref{C5} yields the two-loop approximation
\begin{align}
    \frac{\int_{Y(0)=0}^{Y(T)=0}\mathcal DY\mathcal D\bar b\mathcal Db \exp(iW)}{\int_{Y(0)=0}^{Y(T)=0}\mathcal DY\mathcal D\bar b\mathcal Db \exp(iW_0)} 
    =
     1 + \frac{\beta}{12} R + O(\beta^2),
\end{align}
where we let
\begin{align}
    R := c^{i}{}_{ik}c^{j}{}_{jk} +\frac{1}{2}c^{i}{}_{kj}c^j{}_{li} I^{kl}+ \frac{1}{4}c^{i}{}_{jk}c^{l}{}_{mn}I_{il}I^{jm}I^{kn},
    \label{ricci}
\end{align}
expressed here in terms of the (time-independent) structure constants of \eqref{struc} and the inertia tensor of eq.~\eqref{euler-arnold}. The partition function density \eqref{e63bb} thus reads
\begin{equation}
\label{s29}
z(\beta)
         \overset{\beta \rightarrow 0}{\sim}
         \sqrt{\frac{\det I}{(2\pi\beta)^n}}
         \Big(1 + \frac{\beta}{12} R\Big),
\end{equation}
where the leading term is the one-loop result \eqref{e65bb}. One can show that \eqref{ricci} is the Ricci scalar corresponding to the metric \eqref{metric} (see \cite[lemma 1.1]{Milnor:1976}), so \eqref{s29} agrees with the result normally obtained with heat kernel techniques \cite{Vassilevich:2003xt}.

\subsection{Characters of regular representations}
\label{ssec:character}

The character of a group representation maps any group element $g$ on the trace of the operator that represents it. In the case of the left regular representation \eqref{lereg}, the group action of an element $g = \exp(X)$ is implemented by the operator $e^{i\langle \hat p,X\rangle}$. The corresponding character reads
\begin{align}
    \chi(g) 
    =
    \int_G \dd h_{\text{L}} \left(h\right|e^{i\langle \hat p, X\rangle}\left|h\right) = \operatorname{Vol}_{\text{R}}(G)\left(e\right|e^{i\langle \hat p,X\rangle}\left|e\right),
    \label{chiLRR}
\end{align}
where we used \eqref{e42} to similarly simplify the integral by right-invariance. This is nothing but a partition function \eqref{e41} with a Hamiltonian $\propto\langle\hat p,X\rangle$ which is linear in momenta, so it can be evaluated with path integral methods.

Before turning to path integrals, it should be pointed out that the character of a regular representation is rather trivial---as one may expect from the Peter-Weyl theorem \cite{Williams}. This is because the action of left translations on a perfectly localized state at the identity is $e^{i\langle \hat p,X\rangle}|e) = |\exp(X))$, so \eqref{chiLRR} reduces to
\begin{align}
    \chi(g) = \operatorname{Vol}_{\text{R}}(G)\delta_{\text{L}}(g),
    \label{chi is delta}
\end{align}
where $\delta_{\text{L}}$ is the Dirac distribution at the identity for the left Haar measure. The same result can be found from the Frobenius formula for characters of induced representations (see \i{e.g.}~\cite[sec.~3.4]{OblakThesis}).

Let us now recover \eqref{chi is delta} from path integrals. The idea is to write the rightmost equality of \eqref{chiLRR} as a path integral \eqref{e52b} for the Hamiltonian $\hat H = -\langle \hat p, X\rangle$:
\begin{align}
    \frac{\chi(\exp(X))}{\operatorname{Vol}_{\text{R}}(G)} 
    =
    \sum_{Y_f \in \operatorname{Logs}(e)\cap \mathfrak{h}} \int_{Y(0)=0}^{Y(1) = Y_f} \mathcal DY_{\text{R}}\mathcal D\pi\; \exp\bigg[i\int_0^1 \dd t \Big(\langle \pi, \Omega^{\text{R}}_Y(\dot Y) + X\rangle + \langle \pi, X \rangle\Big)\bigg].
\end{align}
The integral over $\pi$ leads to a functional Dirac distribution, so
\begin{align}
    \frac{\chi(\exp(X))}{\operatorname{Vol}_{\text{R}}(G)} 
    =
    \mathcal N \sum_{Y_f \in \operatorname{Logs}(e)\cap \mathfrak{h}} \int_{Y(0)=0}^{Y(1) = Y_f} \mathcal DY_{\text{R}}\, \delta\big(\Omega^{\text{R}}_Y(\dot Y) + X\big)
    \label{funcdelta}
\end{align}
for some normalization $\mathcal N$. The functional delta fixes $\dot h h^{-1} = -X$, where $h=\exp Y$. The unique solution to this equation is $h(t) = \exp(-tX)$, \i{i.e.}~$Y(t) = -tX$. But this satisfies the final time boundary condition only if $Y_f = -X$, which finally implies
\begin{align}
    \frac{\chi(\exp(X))}{\operatorname{Vol}_{\text{R}}(G)} 
    =
    \mathcal N \sum_{Y_f \in \operatorname{Logs}(e)\cap \mathfrak{h}} \delta(Y_f + X)
    =
    \delta_{\text{L}}(\exp X),
\end{align}
where we used \eqref{inclusion loc} in the second equality and fixed the normalization $\mathcal N$ to match the one in the character \eqref{chi is delta}.


\section*{Acknowledgements}

We are grateful to Ismaël Ahlouche Lahlali, Glenn Barnich, Guillaume Bossard, Sylvain Carrozza, Pierre Delplace, Simon Douaud, Benoit Estienne, Bastien Girault and Paul Roux for support and discussions on related subjects.

\appendix
\section{U(1) path integral}
\label{app:U1}

The path integral formulation of quantum mechanics on U(1) is well known (see \textit{e.g.}~\cite{Schulman}, chapter 23) and crucially relies on the fact that functions on U(1) can be seen as periodic functions on $\mathfrak{u}(1)\cong\mathbb{R}$. It serves as a simple toy model to illustrate the issues of compactness encountered in the more general setup of section \ref{sec:G}. In fact, the U(1) case is more than just a warm-up, since it provides a reference point for the regularization of divergent quantities in the generic case.

Here, we first construct the U(1) path integral, then use it to derive the standard propagator of a free particle.

\subsection{Building the path integral}
\label{sapp:U1decomp}

Let us review the standard derivation of the path integral, starting from the evolution operator of the system. The presentation is tailored for generalization to arbitrary Lie groups.

\paragraph{Position and momentum space.} Let $L^2(\text U(1))$ be the Hilbert space of square-integrable functions on U(1). In terms of the exponential coordinate $x\in ]-\pi,\pi[$, the scalar product of wave functions reads
\begin{equation}
\label{s33}
\left( \psi \middle | \phi \right) = \int_{-\pi}^\pi \dd x \;\bar \psi(x)\phi(x).
\end{equation}
The key ingredient for path integrals is that functions on the circle can be seen as $2\pi$-periodic functions on the real line. More precisely, a state $\left|\psi\right) \in L^2(\text{U}(1))$ can be formally mapped to a state $\ket{\iota \psi}\in L^2(\mathbb R)$ through the U(1) version of the inclusion \eqref{iota}:
\begin{align}
      \ket{ \iota \psi}= \frac{1}{\sqrt{|\mathbb{Z}|}}\left| \psi \right).
\end{align}
The scalar product \eqref{s33} is preserved by this mapping, similarly to \eqref{isomiota}: $\bra{\iota \psi}\ket{ \iota \phi} = \frac{1}{|\mathbb{Z}|}\int_\mathbb{R} \dd x \; \bar \psi(x)\phi(x) = \int_{-\pi}^\pi \dd x \; \bar \psi(x)\phi(x) = \left(\psi\middle | \phi\right)$.

Given any $x\in ]-\pi,\pi[$, the corresponding perfectly localized state $\left| x \right)$ is the function that maps $y\mapsto \delta_{\text{U}(1)}(y-x)$. Perfectly localized states on the real line are denoted $\ket x$, each of which maps $y\mapsto \delta(y-x)$. The inclusion in $L^2(\mathbb{R})$ of a perfectly localized state in $L^2(U(1))$ is the Dirac comb
\begin{align}
      \ket{\iota x} 
      = \frac{1}{\sqrt{|\mathbb{Z}|}}\sum_{n\in \mathbb{Z}} \ket{x+2\pi n},
      \label{e9}
\end{align}
which maps $y\mapsto\frac{1}{\sqrt{|\mathbb{Z}|}}\sum_{n\in \mathbb{Z}} \delta(y-x-2\pi n)$. Finally, the momentum basis on $L^2(\mathbb{R})$ is spanned by plane waves $\ket{p}$, each of which maps $y\mapsto e^{ipy}$, so the identity in $L^2(\mathbb R)$ can be written both as $\text{id} = \int \frac{\dd p}{2\pi}\ket{p}\bra{p}$ in momentum space and $\text{id} = \int \dd x\ket{x}\bra{x}$ in position space, where the integrals run over the whole real line.

\paragraph{Path integration.} Given a Hamiltonian operator $\hat H$ acting on $L^2(U(1))$, our goal is to express the propagator $(x_f|  e^{-i\hat HT}|x_i)$ as a path integral. The starting point is to use \eqref{e9} to write the propagator as a sum of decompactified kernels:
\begin{align}
     (x_f|  e^{-i\hat HT}|x_i)
     =
     \frac{1}{|\mathbb{Z}|}\sum_{n,m\in \mathbb{Z}} \bra{x_f + 2\pi m} e^{-i\hat H T} \ket{x_i + 2\pi n},
     \label{a4}
\end{align}
where the Hamiltonian $\hat H$ is now seen as acting $L^2(\mathbb R)$. Since $\hat H$ has a well defined action on $L^2(\text U(1))$, it maps periodic functions to periodic functions and therefore commutes with discrete translations generated by $e^{2\pi i \hat P}$. Writing $\ket{x+2\pi n} = e^{2\pi i n \hat P}\ket{x}$ and commuting the translation operator with the Hamiltonian yields
\begin{align}
     (x_f|  e^{-i\hat HT}|x_i)
     =
     \frac{1}{|\mathbb{Z}|}\sum_{n,m\in \mathbb{Z}}\bra{x_f + 2\pi (m-n)} e^{-i\hat H T}\ket{x_i} = \sum_{w\in \mathbb{Z}}\bra{x_f + 2\pi w} e^{-i\hat H T} \ket{x_i},
     \label{e12}
\end{align}
where we renamed $w = m-n$ to simplify one of the two sums. Note that the formal $\frac{1}{|\mathbb{Z}|}$ factor simplifies in the process, as in eq.~\eqref{e312}. The notation $w$ is chosen to suggest that $w$ will eventually be a winding number.

Each term on the right-hand side of \eqref{e12} is a propagator on $L^2(\mathbb R)$ and can thus be treated via the standard Feynman derivation. Indeed, repeatedly inserting the identity in both position and momentum spaces leads to
\begin{align}
     \bra {x_f + 2\pi w} e^{-i\hat H T} \ket{x_i} 
     =
     \int \prod_{j=1}^{N-1} \left(\frac{\dd x_j \dd p_j}{2\pi}\right) \frac{\dd p_0}{2\pi} \prod_{k=0}^{N-1} \bra{x_{k+1}}\ket{p_k}\bra{p_k}e^{-i\hat H T/N} \ket{x_k},
     \label{e12b}
\end{align}
where $x_N := x_f + 2\pi w$ and $x_0 := x_i$. This is the Abelian analogue of the discretized path integral \eqref{e38}.

Now using $\bra x \ket{p} = e^{ipx}$ and expanding at large $N$, each factor in the product in \eqref{e12b} can be written as
\begin{align}
      \bra{x_{k+1}}\ket{p_k}\bra{p_k}e^{-i\hat H T/N} \ket{x_k} 
      \sim
      e^{i\frac{T}{N}\left(p_k\frac{x_{k+1}-x_k}{\Delta T}-H(x_k,p_k)\right)}.
      \label{e2.7}
\end{align}
Here the symbol of the Hamiltonian is defined as $H(p,x) := \frac{\bra{p}\hat H\ket{x}}{\bra{p}\ket{x}}$, we let $\Delta T := T/N$, and we neglect all terms of order $O(1/N^2)$. Moreover, assuming that $x_k$ interpolates a differentiable function, one expects $\frac{x_{k+1}-x_k}{\Delta T}$ to converge to $\dot x(kT/N) =: \dot x_k$ in the large $N$ limit. Grouping the product of exponentials \eqref{e2.7} into a single exponential, eq.~\eqref{e12b} becomes
\begin{align}
      \bra {x_f +2\pi w} e^{-i\hat H T} \ket{x_i} \sim \int \prod_{j=1}^{N-1} \left(\frac{\dd x_j \dd p_j}{2\pi}\right) \frac{\dd p_0}{2\pi} e^{i\frac{T}{N}\sum_{k=0}^{N-1}\left(p_k \dot x_k - H(x_k,p_k)\right)}.
      \label{e28b}
\end{align}
The exponent in the integrand of \eqref{e28b} is a Riemann sum that converges to the classical action in the large $N$ limit, leading to the path integral
\begin{align}
      \bra {x_f + 2\pi w} e^{-i\hat H T} \ket{x_i} 
      =
      \int_{x(0) = x_i}^{x(T) = x_f+2\pi w}\mathcal{D}x \mathcal{D}p\; e^{i\int_0^T \dd t \,(p\dot x - H(x,p))},
\label{e2.9}
\end{align}
where the formal measure is the Abelian analogue of eq.~\eqref{pamela}, with $n=1$ and no Jacobian factors. Summing all terms of the form \eqref{e2.9} for different values of $w\in\mathbb{Z}$ leads to the final rewriting of the propagator \eqref{e12},
\begin{align}
    (x_f|  e^{-i\hat HT}|x_i)
    =
    \sum_{w\in \mathbb{Z}}\int_{x(0) = x_i}^{x(T) = x_f + 2\pi w}\mathcal{D}x \mathcal{D}p\; e^{i\int_0^T \dd t \,(p\dot x - H(x,p))}.
      \label{e17}
\end{align}
Note that the sum over $w$ is topological, since $w$ is the winding number of the trajectory $x(t)$ between times 0 and $T$. Put differently, $w$ labels the branch of the logarithm in which one ends up at the end of the motion. It is also clear from the derivation that, at each time step, the integration on $x(t)$ should be done over the whole real line $\mathbb{R}$, and not simply over the $]-\pi,\pi[$ interval. This crucial point ensures that Gaussian perturbations around classical trajectories can be performed without any subtlety on integration bounds; it also explains why perturbation theory is exact for quadratic Hamiltonians on U(1).

\subsection{Euler-Arnold dynamics on U(1); free particle on a circle}
\label{sapp:U1free}

Let us compute the path integral \eqref{e17} in the simplest case of a quadratic Hamiltonian $\hat H = \frac{\hat P^2}{2M}$.\footnote{We use dimensions such that $\hbar=1$ and the position coordinate $x$ in \eqref{s33} is dimensionless, so momentum is dimensionless as well. Time has units of mass, while energy has units $1/\text{mass}$.} Its symbol, defined below \eqref{e2.7}, is just $H(x,p) = \frac{p^2}{2M}$, so the path integral \eqref{e17} can be written as
\begin{align}
       (x_f| e^{-i\hat HT}|x_i)
       = 
       \sum_{w\in \mathbb{Z}}\int_{x(0) = x_i}^{x(T) = x_f + 2\pi w}\mathcal{D}x \mathcal{D}p\; e^{i\int_0^T \dd t \,\left(p\dot x - \frac{p^2}{2M}\right)}.
\label{e211}
\end{align}
The action appearing here is quadratic in $p$, so Gaussian integration over momenta is straightforward:
\begin{align}
       (x_f|  e^{-i\hat HT}|x_i)
       =
       \mathcal N\sum_{w\in \mathbb{Z}}\int_{x(0) = x_i}^{x(T) = x_f + 2\pi w}\mathcal{D}x \; e^{i\int_0^T \dd t \;M\frac{\dot x^2}{2}},
       \label{e13}
\end{align}
where $\mathcal N:=\lim_{N\to\infty}(MN/2\pi i T)^{N/2}$ is a divergent normalization and the formal integration measure is $\mathcal{D}x:=\lim_{N\to\infty}\prod_{j=1}^{N-1}\dd x_j$. The saddle point of the remaining Lagrangian action in \eqref{e13} is the classical trajectory $x_{\text{cl}}(t)$ that solves the equation of motion $\ddot x = 0$ with the boundary conditions $x(0) = x_i$ and $x(T) = x_f + 2\pi w$, that is, $x_{\text{cl}}(t) = x_i + \frac{t}{T}(x_f-x_i+2\pi w)$. Changing variables to $x(t) = x_{\text{cl}}(t) + y(t)$ and using translation-invariance of the integration measure, one finds the factorized propagator
\begin{align}
       (x_f|  e^{-i\hat HT}|x_i)
       =
       \mathcal N
       \left(\int_{y(0) = 0}^{y(T) = 0}\mathcal{D}y \; e^{i\int_0^T \dd t \;M\frac{\dot y^2}{2}}\right)
       \left(\sum_{w\in \mathbb{Z}}e^{i\int_0^T \dd t M\frac{\dot x_{\text{cl}}^2}{2}}\right).
       \label{e20}
\end{align}
All the dependence of this expression on $x_f$ and $x_i$ is contained in the rightmost sum. The latter can be expressed in terms of the Jacobi theta function:
\begin{align}
       \sum_{w\in \mathbb{Z}}
       e^{i\int_0^T \dd t M\frac{\dot x_{\text{cl}}^2}{2}} 
       =
       e^{i\frac{M(x_f-x_i)^2}{2T}}\vartheta\left(M\frac{x_f-x_i}{T} ; \frac{2\pi M}{T}\right)
       \quad\text{with}\quad\vartheta(z;\tau) 
       :=
       \sum_{w\in \mathbb{Z}}e^{i\pi w^2\tau + 2\pi i w z}.
\end{align}
Some time dependence remains inside the middle integral in \eqref{e20}. It is computed using a mode expansion for functions on $[0,T]$ with Dirichlet boundary conditions: $y(t) = \sum_{n=1}^{\infty} a_n[r] \sqrt{\frac{2}{T}}\sin(\frac{n\pi t}{T})$, where the $\sqrt{\frac{2}{T}}$ factor is introduced to make the transformation orthogonal:
\begin{align}
       \frac{2}{T}\int_0^T \sin(\frac{n\pi t}{T})\sin(\frac{m\pi t}{T}) = \delta_{nm}.
       \label{e215}
\end{align}
This ensures that the functional Jacobian is trivial, $\mathcal Dy = \prod_{n=1}^{\infty} \dd a_n$. Furthermore, each mode is an eigenvector of $-\partial_t^2$, so the path integral \eqref{e20} reduces to an infinite product of Gaussian integrals:
\begin{align}
      \int_{y(0) = 0}^{y(T) = 0}\mathcal{D}y \; e^{i\int_0^T \dd t \;M\frac{\dot y^2}{2}}
      = 
      \prod_{n=1}^{\infty} \int \dd a_n\; e^{iM\frac{a_n^2 n^2 \pi^2}{2T^2}}
       \propto 
       \prod_{n=1}^{\infty}\frac{T}{n} 
       =
       e^{\sum_{n=1}^{\infty}\log(\frac{T}{n})}.
       \label{e24}
\end{align}
It only remains to assign a finite value to the divergent sum on the right-hand side of \eqref{e24}. One can do so by resorting to zeta-function regularization,
\begin{equation}
       \sum_{n=1}^{\infty}\log(\frac{T}{n})
       =
       -\frac{\dd}{\dd s}\Big|_{s=0}\sum_{n=1}^\infty \left(\frac{T}{n}\right)^{-s}
       =
       \log(T)\zeta(0) -\zeta'(0)
      = 
      -\frac{1}{2}\log(T)+\frac{1}{2}\log(2\pi).
       \label{e19}
\end{equation}
Plugging this into \eqref{e24} produces the $1/\sqrt{T}$ behaviour of the propagator. Finally, the normalization constant is found by evaluating the kernel in the limit $T\to0$, at which point the propagator reduces to a delta function. This ultimately yields
\begin{align}
       (x_f|e^{-i\hat HT}|x_i)
       =
       \sqrt{\frac{M}{2\pi i T}}e^{iM\frac{(x_f-x_i)^2}{2T}}\vartheta\left(M\frac{x_f-x_i}{T};\frac{2\pi M}{T}\right)
       \label{ea19}
\end{align}
for the U(1) propagator of a free particle. It is directly analogous to the one-loop expansion \eqref{final 1 loop propagator} for an arbitrary Euler-Arnold system. Note a general lesson to be learned from this derivation: up to normalizations that can be determined after the fact, one always has
\begin{align}
        \int_{y(0) = 0}^{y(T) = 0}\mathcal{D}y \; e^{i\int_0^T \dd t \;\frac{\dot y^2}{2}} = \frac{1}{\sqrt{\operatorname{Det}(-\partial_t^2)}} = \frac{1}{\sqrt{2\pi i T}},
        \label{e21}
\end{align}
so that $\operatorname{Det}(-\partial_t^2) = 2\pi i T$. This is a key reference point for the evaluation of one-loop path integrals of Euler-Arnold systems, stated in eq.~\eqref{t23}.

\subsection{Two-point functions}

The Hamiltonian of a free particle is quadratic and the U(1) group is Abelian, so the corresponding path integral is one-loop exact. This provides the template for the computation of two-point functions such as \eqref{e361}. Here, we compute the two-point correlator of the variable $y$ for the Dirichlet-Dirichlet problem on the time interval $[0,T]$.

Recall that, by definition,
\begin{align}
        \langle y(t)y(s)\rangle
        :=
        \frac{\int_{y(0)=0}^{y(T)=0}\mathcal D y\; y(t)y(s) e^{i\int_0^T\dd t \frac{\dot y^2}{2}}}{\int_{y(0)=0}^{y(T)=0}\mathcal D y\; e^{i\int_0^T\dd t \frac{\dot y^2}{2}}}
        =
        \frac{2}{T}\sum_{n=1}^{\infty}\sum_{m=1}^\infty \langle a_n a_m\rangle \sin(\frac{n\pi t}{T})\sin(\frac{m\pi s}{T})
        \label{e31},
\end{align}
where the second equality stems from the mode expansion defined above \eqref{e215}. The correlator $\langle a_na_m\rangle$ can be computed by adding a source term in the path integral expressed in terms of modes \eqref{e24}. Indeed, define
\begin{align}
        Z_J := \int \prod_{k=1}^{\infty}\dd a_k\; e^{i\left(\frac{a_k^2 k^2 \pi^2}{2T^2} + J_ka_k\right)} = \frac{1}{\sqrt{2\pi i T}}\prod_{k=1}^{\infty} e^{-\frac{iJ_k^2T^2}{2\pi^2k^2}},
\end{align}
where the right-hand side is obtained by completing the square. Then the mode correlator is obtained by differentiation:
\begin{align}
    \langle a_n a_m\rangle 
    =
    - \sqrt{2\pi i T}\frac{\partial^2 Z_J}{\partial J_n\partial J_m}\Big|_{J=0} = \frac{T^2}{i\pi^2}\frac{\delta_{nm}}{n^2}.
    \label{e33}
\end{align}
Plugging this two-point function into \eqref{e31} and using the identity $\sum_{n=1}^\infty \frac{\cos(n\tau)}{n^2}= \frac{\pi^2}{6}+\frac{\tau^2}{4}-\frac{\pi|\tau|}{2}$ for $\tau\in [-2\pi,2\pi]$, one obtains the two-point function
\begin{align}
    \langle y(t)y(s)\rangle = i\left(\min(t,s)-\frac{ts}{T}\right).
    \label{e28}
\end{align}
Note that this differs from the standard free scalar propagator. In particular, it is not a function of $t-s$, due to the breaking of translation invariance by Dirichlet boundary conditions. A similar expression is found in \eqref{greenpartition} for expansions around the trivial saddle in the general Lie group case.


\section{Lie group geometry in exponential coordinates}
\label{app:B}

This appendix gathers some geometric properties of Lie groups used in the main text, all expressed in terms of the exponential coordinates defined in section \ref{ssec:ops}.

\paragraph{Maurer-Cartan forms.} Recall that the right and left Maurer-Cartan forms were respectively introduced in eqs.~\eqref{liouville} and \eqref{e312bis}. Let us write their expression in exponential coordinates. 

Consider a curve $g(t) = \exp(X(t))$ in $G$, with velocity
\begin{align}
        \dot g
        =
        \sum_{j=1}^{\infty}\sum_{k=0}^{j-1}\frac{X^k\dot X X^{j-k-1}}{j!}
        =
        \sum_{n=0}^\infty\sum_{m=0}^\infty \frac{X^n\dot X X^m}{(m+n+1)!},
        \label{e68}
\end{align}
where the last equality is obtained by changing indices to $n=k$ and $m=i-k-1$. Now noting that $\int_0^1 \dd s (1-s)^n s^m = \frac{n!m!}{(m+n+1)!}$, this can be rewritten as
\begin{align}
        \dot g = \int_0^1 \dd s \sum_{n=0}^\infty\sum_{m=0}^\infty
        \frac{(1-s)^nX^n}{n!}\dot X \frac{s^mX^m}{m!} = \int_0^1\dd s\; \exp((1-s)X)\dot X \exp(sX).
        \label{ea3}
\end{align}
It follows that the left Maurer-Cartan form at $g=\exp(X)$, acting on the tangent vector \eqref{e68}, reads
\begin{align}
         g^{-1}\dot g 
         = \int_0^1\dd s\;\text{Ad}_{\exp(-sX)}\dot X = \int_0^1\dd s\; e^{-s\text{ad}_{X}}\dot X 
         =
         \frac{1-e^{-\text{ad}_{X}}}{\text{ad}_{X}}\dot X.
\end{align}
A similar argument applies to the right Maurer-Cartan. All in all, in exponential coordinates, the left and right Maurer-Cartan forms read
\begin{align}
        \Omega^{\text{L}}_X(Y) = \frac{1-e^{-\text{ad}_{X}}}{\text{ad}_{X}}Y,
        \qquad
        \Omega^{\text{R}}_X(Y) = \frac{e^{\text{ad}_{X}}-1}{\text{ad}_{X}}Y,
        \label{e71}
\end{align}
where we abuse notation as explained above \eqref{e312bis}: what we write as $\Omega^{L/R}_X$ is technically the pullback (to the Lie algebra $\mathfrak{g}$) of the left/right Maurer-Cartan form $\Omega^{L/R}$ (defined on $G$) by the exponential map.

\paragraph{Haar measures.} Haar measures crucially appear in all integrals over the group $G$ or its Lie algebra $\mathfrak{g}$: the left one beginning in eq.~\eqref{posrepscalar}, the right one in \eqref{e52b}. Let us therefore express them in exponential coordinates. As a by-product, the derivation confirms that left and right invariant measures on $G$ are equivalent, in the sense that they are related by a suitable Jacobian (the modular function \eqref{mofo}).

Recall the construction of invariant metrics sketched around eq.~\eqref{s7}: given an inertia tensor $I:\mathfrak{g}\rightarrow \mathfrak{g}^*$ as in \eqref{euler-arnold}, define a left (right) invariant metric $\gamma$ on G by
\begin{align}
        \gamma_{L(R)}(u,v) 
        := 
        \langle I \Omega^{L(R)}(u), \Omega^{L(R)}(v)\rangle \qquad \text{for all } u,v\in T_gG,
        \label{ea6}
\end{align}
where $\Omega^{L(R)}$ denotes the left (right) Maurer-Cartan form. The density $\sqrt{\det\gamma_{L(R)}}$ defines a left (right) invariant measure on the group. Given that Maurer-Cartan forms in exponential coordinates are given by \eqref{e71}, this density reads
\begin{align}
        \sqrt{\det \gamma_{\text{L}}} 
        &=
        \sqrt{\det I \det(\frac{1-e^{-\text{ad}_{X}}}{\text{ad}_{X}})^2} 
        &\propto~~~
        \det(\frac{1-e^{-\text{ad}_{X}}}{\text{ad}_{X}})
        &=: J_{\text{L}}(X),
        \label{e83bis}
        \\
        \sqrt{\det \gamma_{\text{R}}} &= \sqrt{\det I \det(\frac{e^{\text{ad}_{X}}-1}{\text{ad}_{X}})^2}
        &\propto~~~
        \det(\frac{e^{\text{ad}_{X}}-1}{\text{ad}_{X}})
        &=:J_{\text{R}}(X).
        \label{e83}
\end{align}
Here we dropped the factor $\sqrt{\det I}$ in the definitions of $J_{\text{L}}$ and $J_{\text{R}}$ in order to make them independent of the choice of inertia. The Jacobian that relates a Haar measure to the flat measure is therefore nothing but the determinant of a Maurer-Cartan form.

In general, the left and right Haar measures differ. This can be seen \i{e.g.}~by factorising by $\det(e^{\pm\text{ad}_{X}/2})$ in \eqref{e83}:
\begin{equation}
        J_{\text{R}}(X) = \det(e^{\text{ad}_{X}/2}) \det(\frac{\sinh(\text{ad}_{X/2})}{\text{ad}_{X/2}}),\quad 
       J_{\text{L}}(X) = \det(e^{-\text{ad}_{X}/2}) \det(\frac{\sinh(\text{ad}_{X/2})}{\text{ad}_{X/2}}).
       \label{ea8}
\end{equation}
It follows that the modular function \eqref{mofo} reads
\begin{align}
    \Delta(\exp(X)) := \frac{J_{\text{L}}(X)}{J_{\text{R}}(X)} = \frac{1}{\det(\text{Ad}_{\exp(X)})} = \frac{1}{e^{\tr(\text{ad}_X)}}.
    \label{modular function}
\end{align}
in exponential coordinates. Note that the modular function is multiplicative, since it is expressed in terms of the determinant of a group action: $\Delta(gh)=\Delta(g)\Delta(h)$.

Recall finally that a group is \i{unimodular} if its left and right Haar measures coincide, \i{i.e.}~when its modular function is $\Delta(g)=1$ for all $g\in G$. From eqs.~\eqref{ea8}, this is the case if and only if $\tr(\text{ad}_{X}) = 0$, which can be recast as $c^i{}_{ji}=0$ in terms of the structure constants introduced above \eqref{e22}. For unimodular groups, the Haar measure can be put in the more symmetric form
\begin{align}
        J(X) = \det(\frac{\sinh(\text{ad}_{X/2})}{\text{ad}_{X/2}}).
        \label{e77}
\end{align}

\paragraph{Series expansions.} The expressions \eqref{e71} for the Maurer-Cartan forms and \eqref{e77} for the Jacobian matrix of unimodular groups are essential in perturbation theory. Indeed, they are readily expanded in powers of $X$, with higher-order terms appearing at higher loop orders, as in section \ref{ssecTwoLoop}. Expanding Maurer-Cartan forms \eqref{e71} and the matrix in \eqref{e77} yields
\begin{align}
        \frac{1-e^{-\text{ad}_{X}}}{\text{ad}_{X}}Y &= \sum_{n=0}^{\infty} \frac{(-1)^n}{(n+1)!}\text{ad}_{X}^n Y = Y -\frac{1}{2}[X,Y] + O(X^2),
        \label{e62}
        \\
        \frac{e^{\text{ad}_{X}}-1}{\text{ad}_{X}} Y &= \sum_{n=0}^{\infty} \frac{1}{(n+1)!}\text{ad}_{X}^n Y= Y + \frac{1}{2}[X,Y] + O(X^2),
        \label{e63}
        \\
        \frac{\sinh(\text{ad}_{X/2})}{\text{ad}_{X/2}} Y &=
       \sum_{n=0}^{\infty} \frac{1}{4^n (2n+1)!} \text{ad}_{X}^{2n} Y
       =
       Y + \frac{1}{24}[X,[X,Y]] + O(X^4),
       \label{e64}
\end{align}
where the explicit action of $\text{ad}_Y^n$ in terms of structure constants reads 
\begin{equation}
\left(\text{ad}_{Y}^n\right)^i{}_j = Y^{k_1}\dots Y^{k_n}c^i{}_{k_1l_1}c^{l_1}{}_{k_2l_2}\dots c^{l_{n-2}}{}_{k_{n-1}l_{n-1}}c^{l_{n-1}}{}_{k_n j}.
\end{equation}

\section{Two-loop propagator}
\label{app:2loop}

This appendix presents the complete result of the two-loop computation of propagators for Euler-Arnold Hamiltonians. The result is highly technical but also generic, in the sense that it applies to \textit{any} Euler-Arnold system.

Owing to the explicit expansion of cubic and quartic parts of the action \eqref{e61bb}, respectively in \eqref{W3} and \eqref{W4}, eq.~\eqref{5.39TheHourOfWritingStuff} can be expanded as
\begin{align}
    \frac{\int_{Y(0)=0}^{Y(T)=0}\mathcal DY\mathcal D\bar b\mathcal Db \exp(iW)}{\int_{Y(0)=0}^{Y(T)=0}\mathcal DY\mathcal D\bar b\mathcal Db \exp(iW_0)} = &1 
    + 
    i\int_0^T\dd t\; V^{4Y1D}_{abcd}\langle Y^aY^bY^cY^d\rangle
    +
    i\int_0^T\dd t V^{4Y2D}_{abcd}\langle Y^aY^bY^cY^d\rangle
    \notag
    \\
    & -\frac{1}{2}\int_0^T \dd t\dd s V^{3Y1D}_{abc}V^{3Y1D}_{def}\langle \dot Y^aY^bY^c\dot Y^dY^eY^f\rangle
    \notag
    \\
    & -\frac{1}{2}\int_0^T \dd t\dd s V^{3Y2D}_{abc}V^{3Y2D}_{def}\langle \dot Y^a\dot Y^bY^c\dot Y^d\dot Y^eY^f\rangle
    \notag
    \\
    & -\int_0^T \dd t\dd s V^{3Y2D}_{abc}V^{3Y1D}_{def}\langle \dot Y^a\dot Y^bY^c\dot Y^dY^eY^f\rangle,
    \label{c1}
\end{align}
where tensors with indices $a,b,c$ (resp. $d,e,f$) in the last three lines are understood to be evaluated at point $t$ (resp. $s$), with vertices given by \eqref{3Y1D}--\eqref{3Y2D} and \eqref{4Y1D}--\eqref{4Y2D}, and omitted ghosts. We present the result by treating separately each term in \eqref{c1}. Green's functions are given by \eqref{e361}; equal-time regularization of their derivatives are given by $\partial_1D^{-1}_>(t,t)$ for first derivatives, which can be read off from \eqref{555}, and by the non-diverging part of \eqref{554} for second derivatives. Finally, Lie algebra indices are lowered and raised with $I_{\text{cl}}(t)$ and its inverse, and $^{-1}$ exponents attached to $D$ or $I_{\text{cl}}$ are to be inferred from the position of indices.

We present the raw result of the two-loop computation, but its full time dependenceis known from \eqref{dewitt}. The coefficient $b_2$ of the De Witt expansion \eqref{dewitt} can be inferred from the result of this appendix by replacing $T$ by 1, and all quantities by their reference counterparts.

\paragraph{4Y1D term:}
\begin{align}
    i\int_0^T\dd t\; V^{4Y1D}_{abcd}\langle Y^aY^bY^cY^d\rangle = \frac{i}{8}c^k{}_{a(b}c^j{}_{c|k|}c^{i}{}_{d)j}Q_i\int_0^T\dd t\;\partial_1D^{ab}D^{cd}.
\end{align}
\paragraph{4Y2D term:}
\begin{multline}
    i\int_0^T\dd t\; V^{4Y1D}_{abcd}\langle Y^aY^bY^cY^d\rangle = \int_0^T\dd t\left(\frac{i}{8}c^{i}{}_{ca}c_{ibd}-\frac{i}{6}c_{acj}c^{j}{}_{db}\right)\times\\
    \left(\partial_1\partial_2D^{ab}_{\text{reg}}D^{cd}+\partial_1D^{ac}\partial_1D^{bd}+\partial_1D^{ad}\partial_1D^{bc}\right).
    \label{C3}
\end{multline}
\paragraph{3Y1D--3Y1D term:}
\begin{multline}
    -\frac{1}{2}\int_0^T \dd t\dd s V^{3Y1D}_{abc}V^{3Y1D}_{def}\langle \dot Y^aY^bY^c\dot Y^dY^eY^f\rangle = \frac{i}{72} c^j{}_{a(b}c^i{}_{c)j}Q_ic^k{}_{d(e}c^{l}{}_{f)k}Q_l\times\\
    \int_0^T\dd t\dd s\; \big(4\partial_1D^{ab}\partial_1D^{de}D^{cf}+4\partial_1 D^{ab}\partial_1D^{dc}D^{ef} + 4\partial_1D^{ae}\partial_1D^{db}D^{cf}\\
    + \partial_1\partial_2D^{ad}D^{bc}D^{ef} + 2\partial_1\partial_2D^{ad}D^{be}D^{cf}\big),
\end{multline}
where the times of evaluation of Green functions is inferred from the Lie algebra indices they carry, \textit{e.g.}~$D^{cf}=D^{-1}(t,s)^{cf}$, $\partial_1D^{dc} = \partial_1D(s,t)^{dc}$, etc.
\paragraph{3Y2D-3Y2D term:}
\begin{multline}
    -\frac{1}{2}\int_0^T \dd t\dd s V^{3Y2D}_{abc}V^{3Y2D}_{def}\langle \dot Y^a\dot Y^bY^c\dot Y^d\dot Y^eY^f\rangle = \frac{i}{8}\int_0^T \dd t\dd s \; c_{(ab)c}c_{(de)f}\times\\
    \bigg(\partial_1\partial_2D^{ab}_{\text{reg}}\partial_1\partial_2D^{de}_{\text{reg}}D^{cf}+4\partial_1\partial_2D^{ab}_{\text{reg}}\partial_1D^{dc}\partial_1D^{ef} \\
    + 4\delta(t-s)I_{\text{cl}}^{be}\partial_1\partial_2D^{ad}_{\text{reg}}D^{cf}+ 2\partial_1\partial_2D^{ad}_{\text{reg}}\partial_1\partial_2D^{be}_{\text{reg}}D^{cf}
    \\
    + 4 \partial_1\partial_2D^{ad}\partial_1D^{bf}\partial_1D^{ec}+4 \partial_1\partial_2D^{ad}\partial_1D^{bc}\partial_1D^{ef} \bigg).
    \label{C5}
\end{multline}
\paragraph{3Y2D--3Y1D term:}
\begin{multline}
    -\int_0^T \dd t\dd s V^{3Y2D}_{abc}V^{3Y1D}_{def}\langle \dot Y^a\dot Y^bY^c\dot Y^dY^eY^f\rangle = \frac{i}{12}\int_0^T\dd t \dd s\; c_{(ab)c}c^j{}_{d(e}c^i{}_{f)j}Q_i \times\\
    \bigg(\partial_1\partial_2D^{ab}_{\text{reg}}\partial_1 D^{dc}D^{ef}+2\partial_1\partial_2D^{ab}_{\text{reg}}D^{ce}\partial_1D^{df}+4\partial_1\partial_2D^{ad}D^{ce}\partial_1D^{bf} \\+ 4\partial_1D^{ac}\partial_1D^{de}\partial_1D^{bf}
    +2\partial_1\partial_2D^{ad}\partial_1D^{bc}D^{ef}+2\partial_1D^{ae}\partial_1D^{bf}\partial_1D^{dc}\bigg).
\end{multline}

\providecommand{\href}[2]{#2}

\begingroup\begin{thebibliography}{10}
\addcontentsline{toc}{section}{References}

\bibitem{Beauvillain:2025ygx}
M.~Beauvillain, B.~Oblak, and M.~Petropoulos, ``{Quantum mechanics on Lie
  groups: I. Noncommutative Fourier transforms},'' {\it J. Phys. A} \textbf{59}
  (2026), no.~25, 255303,
  \href{http://www.arXiv.org/abs/2512.19840}{\texttt{2512.19840}}.

\bibitem{Montgomery}
R.~Montgomery, ``{How much does the rigid body rotate? A Berry’s phase from
  the 18th century},'' {\it Amer.~J.~Phys.} \textbf{59} (1991), no.~5,
  394--398.

\bibitem{Marsden}
J.~E. Marsden and T.~S. Ratiu, {\it Introduction to mechanics and symmetry: a
  basic exposition of classical mechanical systems}, vol.~17.
\newblock Springer Science \& Business Media, 2013.

\bibitem{holm2024}
D.~D. Holm, ``{31 Lectures on Geometric Mechanics},''
  \href{http://www.arXiv.org/abs/2408.09564}{\texttt{2408.09564}}.

\bibitem{ArnoldKhesin}
V.~I. Arnold and B.~A. Khesin, {\it Topological methods in hydrodynamics},
  vol.~125.
\newblock Springer Science \& Business Media, 1999.

\bibitem{Khesin}
B.~Khesin and R.~Wendt, {\it {The geometry of infinite-dimensional groups}},
  vol.~51.
\newblock Springer Science \& Business Media, 2008.

\bibitem{ArnoldOrigin}
V.~I. Arnold, ``{Sur la g{\'e}om{\'e}trie diff{\'e}rentielle des groupes de Lie
  de dimension infinie et ses applications \`a l'hydrodynamique des fluides
  parfaits},'' {\it Ann. Inst. Fourier} \textbf{16} (1966), no.~1, 319--361.

\bibitem{Vasylkevych}
S.~Vasylkevych and J.~E. Marsden, ``{The Lie-Poisson Structure of the Euler
  Equations of an Ideal Fluid},'' {\it Dyn.~PDE} \textbf{2} (2005), no.~4,
  281--300, \href{http://www.arXiv.org/abs/0711.4875}{\texttt{0711.4875}}.

\bibitem{Modin}
K.~Modin, ``{Geometric Hydrodynamics: from Euler, to Poincar\'e, to Arnold},''
  \href{http://www.arXiv.org/abs/1910.03301}{\texttt{1910.03301}}.

\bibitem{holm1998euler}
D.~D. Holm, J.~E. Marsden, and T.~S. Ratiu, ``{The Euler--Poincar{\'e}
  equations and semidirect products with applications to continuum theories},''
  {\it Adv.~Math.} \textbf{137} (1998), no.~1, 1--81,
  \href{http://www.arXiv.org/abs/chao-dyn/9801015}{\texttt{chao-dyn/9801015}}.

\bibitem{holm1999euler}
D.~D. Holm, J.~E. Marsden, and T.~S. Ratiu, ``{The Euler-Poincar{\'e} equations
  in geophysical fluid dynamics},''
  \href{http://www.arXiv.org/abs/chao-dyn/9903035}{\texttt{chao-dyn/9903035}}.

\bibitem{gaybalmaz2009reduced}
F.~Gay-Balmaz and T.~S. Ratiu, ``{Reduced Lagrangian and Hamiltonian
  formulations of Euler-Yang-Mills fluids},''
  \href{http://www.arXiv.org/abs/0903.4287}{\texttt{0903.4287}}.

\bibitem{OblakKozy}
B.~Oblak and G.~Kozyreff, ``{Berry Phases in the Reconstructed KdV Equation},''
  {\it Chaos} \textbf{30} (2020) 113114,
  \href{http://www.arXiv.org/abs/2002.01780}{\texttt{2002.01780}}.

\bibitem{Gripaios:2014yha}
B.~Gripaios and D.~Sutherland, ``{Quantum Field Theory of Fluids},'' {\it Phys.
  Rev. Lett.} \textbf{114} (2015), no.~7, 071601,
  \href{http://www.arXiv.org/abs/1406.4422}{\texttt{1406.4422}}.

\bibitem{Endlich:2010hf}
S.~Endlich, A.~Nicolis, R.~Rattazzi, and J.~Wang, ``{The Quantum mechanics of
  perfect fluids},'' {\it JHEP} \textbf{04} (2011) 102,
  \href{http://www.arXiv.org/abs/1011.6396}{\texttt{1011.6396}}.

\bibitem{Dersy:2022kjd}
A.~Dersy, A.~Khmelnitsky, and R.~Rattazzi, ``{The quantum perfect fluid in
  2D},'' {\it SciPost Phys.} \textbf{17} (2024), no.~1, 019,
  \href{http://www.arXiv.org/abs/2211.09820}{\texttt{2211.09820}}.

\bibitem{LandauLifschitz}
L.~Landau and E.~Lifshitz, ``On the theory of the dispersion of magnetic
  permeability in ferromagnetic bodies,'' {\it Phys. Zeitsch. der Sow.}
  \textbf{8} (1935) 153--169.

\bibitem{Lakshmanan}
M.~Lakshmanan, ``{The fascinating world of the Landau--Lifshitz--Gilbert
  equation: an overview},'' {\it Philos.~Trans.~Roy.~Soc.~A} \textbf{369}
  (2011), no.~1939, 1280--1300,
  \href{http://www.arXiv.org/abs/1101.1005}{\texttt{1101.1005}}.

\bibitem{Jevicki:1978yv}
A.~Jevicki and N.~Papanicolaou, ``{Semiclassical Spectrum of the Continuous
  Heisenberg Spin Chain},'' {\it Annals Phys.} \textbf{120} (1979) 107.

\bibitem{gaudin2014}
M.~Gaudin, {\it The Bethe wavefunction}.
\newblock Cambridge University Press, 2014.

\bibitem{Lamers}
J.~Lamers, ``{A pedagogical introduction to quantum integrability, with a view
  towards theoretical high-energy physics},'' {\it PoS} \textbf{Modave2014}
  (2015) 001, \href{http://www.arXiv.org/abs/1501.06805}{\texttt{1501.06805}}.

\bibitem{Nussle:2025umz}
T.~Nussle, S.~Nicolis, I.~Sofos, and J.~Barker, ``{Path-integral spin dynamics
  with exchange and external field},'' {\it Phys. Rev. B} \textbf{112} (2025),
  no.~5, 054404,
  \href{http://www.arXiv.org/abs/2502.19113}{\texttt{2502.19113}}.

\bibitem{Delacretaz}
L.~V. Delacretaz, Y.-H. Du, U.~Mehta, and D.~T. Son, ``{Nonlinear bosonization
  of Fermi surfaces: The method of coadjoint orbits},'' {\it Phys. Rev. Res.}
  \textbf{4} (2022), no.~3, 033131,
  \href{http://www.arXiv.org/abs/2203.05004}{\texttt{2203.05004}}.

\bibitem{Huang:2023hbt}
X.~Huang, ``{Effective field theory of Berry Fermi liquid from the coadjoint
  orbit method},'' {\it Phys. Rev. B} \textbf{109} (2024), no.~23, 235146,
  \href{http://www.arXiv.org/abs/2312.00877}{\texttt{2312.00877}}.

\bibitem{Beauvillain:2024dou}
M.~Beauvillain, B.~Oblak, and M.~Petropoulos, ``{Berry phases in the
  bosonization of nonlinear edge modes},'' {\it Phys. Rev. B} \textbf{112}
  (2025), no.~12, 125136,
  \href{http://www.arXiv.org/abs/2408.03991}{\texttt{2408.03991}}.

\bibitem{Allen}
H.~C. Allen and P.~C. Cross, {\it Molecular Vib-rotors: The theory and
  interpretation of high resolution infrared spectra}.
\newblock Wiley, 1963.

\bibitem{Bauder}
A.~Bauder, {\it Fundamentals of rotational spectroscopy}.
\newblock Wiley Online Library, 2011.

\bibitem{Biedenharn:1984}
L.~C. Biedenharn and J.~D. Louck, {\it Angular Momentum in Quantum Physics:
  Theory and Application}.
\newblock Encyclopedia of Mathematics and its Applications. Cambridge
  University Press, 1984.

\bibitem{Gripaios:2015pfa}
B.~Gripaios and D.~Sutherland, ``{Quantum mechanics of a generalised rigid
  body},'' {\it J. Phys. A} \textbf{49} (2016), no.~19, 195201,
  \href{http://www.arXiv.org/abs/1504.01406}{\texttt{1504.01406}}.

\bibitem{Freidel:2005bb}
L.~Freidel and E.~R. Livine, ``{Ponzano-Regge model revisited III: Feynman
  diagrams and effective field theory},'' {\it Class. Quant. Grav.} \textbf{23}
  (2006) 2021--2062,
  \href{http://www.arXiv.org/abs/hep-th/0502106}{\texttt{hep-th/0502106}}.

\bibitem{Freidel:2005me}
L.~Freidel and E.~R. Livine, ``{3D Quantum Gravity and Effective Noncommutative
  Quantum Field Theory},'' {\it Phys. Rev. Lett.} \textbf{96} (2006) 221301,
  \href{http://www.arXiv.org/abs/hep-th/0512113}{\texttt{hep-th/0512113}}.

\bibitem{Freidel:2005ec}
L.~Freidel and S.~Majid, ``{Noncommutative harmonic analysis, sampling theory
  and the Duflo map in 2+1 quantum gravity},'' {\it Class. Quant. Grav.}
  \textbf{25} (2008) 045006,
  \href{http://www.arXiv.org/abs/hep-th/0601004}{\texttt{hep-th/0601004}}.

\bibitem{Joung:2008mr}
E.~Joung, J.~Mourad, and K.~Noui, ``{Three Dimensional Quantum Geometry and
  Deformed Poincare Symmetry},'' {\it J. Math. Phys.} \textbf{50} (2009)
  052503, \href{http://www.arXiv.org/abs/0806.4121}{\texttt{0806.4121}}.

\bibitem{Dupuis:2011fx}
M.~Dupuis, F.~Girelli, and E.~R. Livine, ``{Spinors and Voros star-product for
  Group Field Theory: First Contact},'' {\it Phys. Rev. D} \textbf{86} (2012)
  105034, \href{http://www.arXiv.org/abs/1107.5693}{\texttt{1107.5693}}.

\bibitem{Oriti:2011ac}
D.~Oriti and M.~Raasakka, ``{Quantum Mechanics on SO(3) via Non-commutative
  Dual Variables},'' {\it Phys. Rev. D} \textbf{84} (2011) 025003,
  \href{http://www.arXiv.org/abs/1103.2098}{\texttt{1103.2098}}.

\bibitem{Raasakka:2011np}
M.~Raasakka, ``{Group Fourier transform and the phase space path integral for
  finite dimensional Lie groups},''
  \href{http://www.arXiv.org/abs/1111.6481}{\texttt{1111.6481}}.

\bibitem{Guedes:2013vi}
C.~Guedes, D.~Oriti, and M.~Raasakka, ``{Quantization maps, algebra
  representation and non-commutative Fourier transform for Lie groups},'' {\it
  J. Math. Phys.} \textbf{54} (2013) 083508,
  \href{http://www.arXiv.org/abs/1301.7750}{\texttt{1301.7750}}.

\bibitem{Schulman}
Schulman, {\it Techniques and Applications of Path Integration}.
\newblock Dover Publications, 2012.

\bibitem{Mori:2019}
S.~Mori, ``The heat kernel on $sl(2,\mathbb{R})$,'' 2019.

\bibitem{Craddock:2017}
M.~Craddock, ``Fundamental solutions for the two dimensional affine group and
  hn+1,'' {\it Journal of Mathematical Analysis and Applications} \textbf{445}
  (2017), no.~1, 953--970.

\bibitem{David:2010}
J.~R. David, M.~R. Gaberdiel, and R.~Gopakumar, ``The heat kernel on ads 3 and
  its applications,'' {\it Journal of High Energy Physics} \textbf{2010} (Apr.,
  2010).

\bibitem{avramidi:2006}
I.~G. Avramidi, ``Heat kernel asymptotics on symmetric spaces,'' 2006.

\bibitem{Vassilevich:2003xt}
D.~V. Vassilevich, ``{Heat kernel expansion: User's manual},'' {\it Phys.
  Rept.} \textbf{388} (2003) 279--360,
  \href{http://www.arXiv.org/abs/hep-th/0306138}{\texttt{hep-th/0306138}}.

\bibitem{kapranov}
M.~Kapranov, ``Noncommutative geometry and path integrals,''
  \href{http://www.arXiv.org/abs/math/0612411}{\texttt{math/0612411}}.

\bibitem{woodhouse1997}
N.~Woodhouse, {\it {Geometric Quantization}}.
\newblock Oxford mathematical monographs. Clarendon Press, 1997.

\bibitem{Witten:1983ar}
E.~Witten, ``{Nonabelian Bosonization in Two-Dimensions},'' {\it Commun. Math.
  Phys.} \textbf{92} (1984) 455--472.

\bibitem{Gutt}
S.~Gutt, ``An explicit *-product on the cotangent bundle of a lie group,'' {\it
  Letters in Mathematical Physics} \textbf{7} (1983) 249--258.

\bibitem{Dito}
G.~Dito, ``{Kontsevich star-product on the dual of a Lie algebra},'' {\it Lett.
  Math. Phys.} \textbf{48} (1999) 307–322,
  \href{http://www.arXiv.org/abs/math/9905080}{\texttt{math/9905080}}.

\bibitem{Kathotia}
V.~Kathotia, ``{Kontsevich's universal formula for deformation quantization and
  the Campbell--Baker--Hausdorff formula},'' {\it International Journal of
  Mathematics} \textbf{11} (2000), no.~04, 523--551,
  \href{http://www.arXiv.org/abs/math/9811174}{\texttt{math/9811174}}.

\bibitem{Gelfand:1959nq}
I.~M. Gelfand and A.~M. Yaglom, ``{Integration in functional spaces and it
  applications in quantum physics},'' {\it J. Math. Phys.} \textbf{1} (1960)
  48.

\bibitem{Forman:1987gha}
R.~Forman, ``{Functional determinants and geometry},'' {\it Invent. Math.}
  \textbf{88} (1987), no.~3, 447--493.

\bibitem{DeWitt-Morette}
C.~DeWitt-Morette, ``The semiclassical expansion,'' {\it Annals of Physics}
  \textbf{97} (1976), no.~2, 367--399.

\bibitem{Milnor:1976}
J.~Milnor, ``Curvatures of left invariant metrics on lie groups,'' {\it
  Advances in Mathematics} \textbf{21} (1976), no.~3, 293--329.

\bibitem{Williams}
D.~P. Williams, ``{The Peter-Weyl Theorem for Compact Groups},'' {\it Lecture
  notes at Dartmouth College} (1991).

\bibitem{OblakThesis}
B.~Oblak, {\it {BMS Particles in Three Dimensions}}.
\newblock {Springer Theses},
2017.
\newblock

\end{thebibliography}\endgroup
\end{document}